\documentclass[fleqn,usenatbib]{mnras}

\usepackage{newtxtext,newtxmath}
\usepackage{orcidlink}
\usepackage{verbatim}

\usepackage[T1]{fontenc}
\usepackage[utf8]{inputenc}
\DeclareRobustCommand{\VAN}[3]{#2}
\let\VANthebibliography\thebibliography
\def\thebibliography{\DeclareRobustCommand{\VAN}[3]{##3}\VANthebibliography}

\usepackage{graphicx}	% Including figure files
\usepackage{amsmath, mathtools, nccmath, nicefrac}	% amssymb, Advanced maths commands 
\usepackage{gensymb} %for degree symbol
\usepackage{enumitem}
\usepackage{multirow}
\usepackage{textcomp}
\usepackage{chngpage}
\usepackage{footmisc}
\usepackage{bm}
\usepackage{xspace}
\usepackage{tablefootnote, threeparttable}
\usepackage{subfigure}
\usepackage{subcaption}
\usepackage{pdflscape}
\usepackage{float}
\usepackage{makecell}
\usepackage[makeroom]{cancel}
\usepackage{CJK}
\usepackage{arydshln} %for horizontal dashed lines
\usepackage{array}% http://ctan.org/pkg/array

\PassOptionsToPackage{usenames, dvipsnames}{color}
\usepackage{hyperref}
\definecolor{dblue}{RGB}{40, 116, 166}
\definecolor{navy}{RGB}{56, 70, 184}
\definecolor{darkblue}{RGB}{39, 76, 119}
\definecolor{grey}{RGB}{192,192,192}
\definecolor{new_color}{HTML}{CF0000} 
\definecolor{crimson}{RGB}{220,20,60}
\hypersetup{
    colorlinks = true,
    linkcolor = black,
    citecolor= dblue,
    urlcolor = cyan,
}

\newcommand{\masy}{mas\,y$^{-1}$}
\newcommand{\Msun}{\mbox{$M_{\odot}$}}

\newcommand{\Rearth}{R$_{\oplus}$}

\newcommand{\cecilia}{\texttt{cecilia}}

\newcommand{\Gaia}{\emph{Gaia}}
\newcommand{\Galex}{\emph{GALEX}}
\newcommand{\panstarrs}{\emph{Pan-STARRS}}
\newcommand{\SDSS}{\emph{SDSS}}
\newcommand{\KeckESI}{\emph{Keck/ESI}}

\newcommand{\wdzerotwothreeone}{SDSS~J0231$+$2512}
\newcommand{\wdzeroeightfivenine}{SDSS~J0859$+$5732}
\newcommand{\wdoneonezeronine}{SDSS~J1109$+$1318}
\newcommand{\wdonethreethreethree}{SDSS~J1333$+$6364}
\newcommand{\wdtwothreeoneone}{SDSS~J2311$-$0041}

\newcommand{\angs}{\text{\normalfont\AA}}

\newcommand{\Mcvz}{$M_{\rm cvz}$}
\newcommand{\tauZ}{$\tau_{\rm Z}$} 
 
\newcommand{\tauB}{$\tau_{\rm B}$} 
\newcommand{\nA}{$n_{\rm A}$} 
\newcommand{\nB}{$n_{\rm B}$}
\newcommand{\sigmastat}{$\sigma_{\rm{stat, MCMC}}$}
\newcommand{\sigmasys}{$\sigma_{\rm{sys,\texttt{ML}}}$}
\newcommand{\sigmatot}{$\sigma_{\rm{tot}}$}
\newcommand{\cout}[1]{}

\newcommand{\sigmathresh}{$\sigma_{\rm detection}=0.10$~dex}
\newcommand{\sigmafloor}{$\sigma_{\rm floor}=0.20$~dex}

\newcommand{\teff}{$T_{\rm eff}$}                 %1
\newcommand{\logg}{$\log\rm g$}                   %2
\newcommand{\logHHe}{$\log_{10}\rm{(H/He)}$}      %3
\newcommand{\logCaHe}{$\log_{10}\rm{(Ca/He)}$}    %4
\newcommand{\logMgHe}{$\log_{10}\rm{(Mg/He)}$}    %5
\newcommand{\logFeHe}{$\log_{10}\rm{(Fe/He)}$}    %6
\newcommand{\logOHe}{$\log_{10}\rm{(O/He)}$}      %7
\newcommand{\logSiHe}{$\log_{10}\rm{(Si/He)}$}    %8
\newcommand{\logTiHe}{$\log_{10}\rm{(Ti/He)}$}    %9
\newcommand{\logBeHe}{$\log_{10}\rm{(Be/He)}$}    %10
\newcommand{\logCrHe}{$\log_{10}\rm{(Cr/He)}$}    %11
\newcommand{\logMnHe}{$\log_{10}\rm{(Mn/He)}$}    %12
\newcommand{\logNiHe}{$\log_{10}\rm{(Ni/He)}$}    %13
\newcommand{\AjitSDSS}{$\rm{A_{jit, SDSS}}$}  
\newcommand{\AjitKeck}{$\rm{A_{jit, Keck/ESI}}$}  
\newcommand{\RVSDSS}{$\rm{RV_{SDSS}}$}  
\newcommand{\RVKeck}{$\rm{RV_{Keck/ESI}}$}

\title[Compositions of Five White Dwarf Pollutants]{A Machine-Learning Compositional Study of Exoplanetary Material Accreted Onto Five Helium-Atmosphere White Dwarfs with \cecilia}

\author[Badenas-Agusti et al.]{Mariona Badenas-Agusti,$^{1,2,3\thanks{\textrm{E-mail: mbadenas@mit.edu}}}$\orcidlink{0000-0003-4903-567X} 
Siyi Xu \begin{CJK*}{UTF8}{gbsn}(许\CJKfamily{bsmi}偲\CJKfamily{gbsn}艺\end{CJK*}),$^{4}$\orcidlink{0000-0002-8808-4282}
Andrew Vanderburg,$^{2}$\orcidlink{0000-0001-7246-5438}
Kishalay De,$^{2}$\newauthor
Patrick Dufour,$^{5}$\orcidlink{0000-0003-4609-4500}
Laura K. Rogers,$^{1,4}$\orcidlink{https://orcid.org/0000-0002-3553-9474}
Susana Hoyos,$^{2,6}$\orcidlink{0000-0001-9808-027X} 
Simon Blouin,$^{7}$\orcidlink{0000-0002-9632-1436}
Javier Via$\tilde{\rm{n}}$a,$^{2}$\orcidlink{0000-0002-0563-784X}\newauthor
Amy Bonsor,$^{1}$\orcidlink{0000-0002-8070-1901}
Ben Zuckerman$^{8}$\orcidlink{https://orcid.org/0000-0001-6809-3045}\\
$^{1}$Institute of Astronomy, University of Cambridge, Madingley Road, Cambridge, CB3 0HA, UK\\
$^{2}$Department of Earth, Atmospheric and Planetary Sciences, Massachusetts Institute of Technology, Cambridge, MA 02139, USA\\
$^{3}$Department of Physics and Kavli Institute for Astrophysics and Space Research, Massachusetts Institute of Technology, Cambridge, MA 02139, USA\\
$^{4}$Gemini Observatory/NSF's NOIRLab, 950 N. Cherry Ave, Tucson, AZ, 85719, USA  \\
$^{5}$D\'epartement de Physique, Universit\'e de Montr\'eal, Montr\'eal, Qu\'ebec H3C 3J7, Canada \\
$^{6}$Department of Earth, Planetary, and Space Sciences, University of California, Los Angeles, CA 90095-1567, USA\\
$^{7}$Department of Physics and Astronomy, University of Victoria, Victoria, BC V8W 2Y2, Canada\\
$^{8}$Department of Physics and Astronomy, University of California, Los Angeles, CA 90095-1562, USA\\
}

\date{Accepted 2025 May 9. Received 2025 May 7; in original form 2025 January 16}

\pubyear{2025}

\begin{document}
\label{firstpage}
\pagerange{\pageref{firstpage}--\pageref{lastpage}}
\maketitle

% Abstract of the paper
\begin{abstract}
We present the first application of the Machine Learning (ML) pipeline \cecilia{} to determine the physical parameters and photospheric composition of five metal-polluted He-atmosphere white dwarfs without well-characterised elemental abundances. To achieve this, we perform a joint and iterative Bayesian fit to their \SDSS{} (R=2,000) and \KeckESI{} (R=4,500) optical spectra, covering the wavelength range from about 3,800~\angs{} to 9,000~\angs. Our analysis measures the abundances of at least two ---and up to six--- chemical elements in their atmospheres with a predictive accuracy similar to that of conventional WD analysis techniques ($\approx$0.20~dex). The white dwarfs with the largest number of detected heavy elements are \wdzeroeightfivenine{} and \wdtwothreeoneone{}, which simultaneously exhibit O, Mg, Si, Ca, and Fe in their \KeckESI{} spectra. For all systems, we find that the bulk composition of their pollutants is largely consistent with those of primitive CI chondrites to within 1-2$\sigma$. We also find evidence of statistically significant ($>2\sigma$) oxygen excesses for \wdzeroeightfivenine{} and \wdtwothreeoneone{}, which could point to the accretion of oxygen-rich exoplanetary material. In the future, as wide-field astronomical surveys deliver millions of public WD spectra to the scientific community, \cecilia{} aspires to unlock population-wide studies of polluted WDs, therefore helping to improve our statistical knowledge of extrasolar compositions.

\end{abstract}

% Select between one and six entries from the list of approved keywords.
% Don't make up new ones.
%%%%%%%%%%%%%%%%%%%%%%%%%%%%%%%%%%%%%%%%%%%%%%%%%%

%%%%%%%%%%%%%%%%% BODY OF PAPER %%%%%%%%%%%%%%%%%%

\section{Introduction}

For several decades, the search and characterisation of exoplanets has primarily relied on two indirect observational techniques: transit photometry, which measures periodic changes in the flux of a star due to a transiting planet \citep{Charbonneau:2000}, and high-precision radial velocities, which track the Doppler shift of a star's spectral lines induced by a planet's gravitational pull \citep{Campbell:1988}. These techniques provide, respectively, the radius and mass of the planet, which can then be combined to estimate its bulk density and infer its most plausible interior composition. At the heart of this characterisation process are theoretical Mass-Radius (MR) diagrams in which the planet's radius and mass are compared to synthetic interior models \citep{Seager:2007, Fortney:2007, Zeng:2008, Zeng:2013, Dressing:2015}. These diagrams have proved useful to broadly differentiate between rocky or gaseous planets \citep{Zeng:2016}. However, they are strongly degenerate as they can yield more than one possible core composition for an identical mass and radius measurement \citep{Rogers:2010, Dorn:2015}. This ambiguity is further complicated by the limited number of exoplanets with well-measured masses and radii \citep{Jontof:2019}, which makes it difficult to distinguish between different planetary interiors.

As we seek to improve our knowledge of extrasolar bodies, it will be crucial to mitigate the degeneracies of conventional exoplanet characterisation techniques. Fortunately, spectroscopic observations of ``polluted'' White Dwarfs (WDs) provide a solution to this problem by facilitating accurate measurements of the bulk composition of exoplanetary material \citep[e.g.][]{Zuckerman:2007, Klein:2010, Gansicke:2012, Rogers:2024a}. WDs are the degenerate cooling remnants of low- and intermediate-mass ($\leq8$~\Msun) Main-Sequence (MS) stars \citep{WeidemannKoester:1983}. Their radius is similar to that of the Earth  ($R_{\rm{WD}}$$\approx$1~\Rearth), but their mass is half that of the Sun  ($M_{\rm{WD}}$$\approx$0.6~\Msun). As a result, they are extremely dense and compact, with surface gravities on the order 10$^{8}~\rm{cm/s^2}$. Due to their strong gravitational fields, WDs should have pristine atmospheres composed only of lightweight elements (i.e. H and/or He), with elements heavier than He (or ``metals;'' atomic number Z$>$2) sinking rapidly towards the unobservable stellar interior in timescales much shorter than the evolutionary age of the star \citep{Fontaine:1979, Paquette:1986}.\footnote{Diffusion timescales of heavy elements can vary from days to thousands of years in warm, H-atmosphere WDs (spectral type DA), to millions of years in cool, He-dominated systems (spectral type DB and DZ) \citep{Koester:2009}.} Nevertheless, contrary to this expectation, observations suggest that between 25$\%$ to 50$\%$ of isolated WDs are contaminated with traces of metals, such as calcium (Ca), magnesium (Mg), oxygen (O), silicon (Si), or iron (Fe) \citep[e.g.][]{Zuckerman:2003, Zuckerman:2010, Koester:2014}. At temperatures lower than \teff$\lesssim$25,000~K, this phenomenon is likely due to the recent or ongoing accretion of tidally disrupted material from a planetary system that survived the post-MS evolution of its host star \citep[e.g. see reviews by][]{JuraYoung:2014, Farihi:2016, Veras:2021, Xu:2024_review}. In hotter systems, the outward flow of radiative levitation pressure can also bring heavy elements to the surface \citep{Chayer:1995}, while in cool (\teff$\lesssim$10,000~K), helium-dominated WDs, the presence of carbon can be attributed to convective dredge-up from the deep stellar interior \citep{Pelletier:1986, Camisassa:2017, Bedard:2022b}.

Since the discovery of three metals in the atmosphere of the He-rich white dwarf ``van Maanen 2'' \citep{VanMaanen:1917, Weidemann:1960}, the field of polluted WDs has consolidated into a valuable discipline to infer the geology and chemistry of extrasolar bodies. This inference process typically consists of two stages: first, detailed atmosphere models are used to fit a WD spectrum and obtain the star's photospheric abundances \citep[e.g.][]{Koester:2009, Koester:2010, Dufour:2007}. Next, the observed stellar abundances are used to constrain the bulk composition of the polluting body via WD accretion and diffusion equations \citep{Koester:2009, JuraYoung:2014}. In general, the most polluted WDs only exhibit one or two metals in their spectra \citep{Williams:2024}, especially Ca and Mg in the optical \citep{Coutu:2019}, and Si in the Ultraviolet (UV) \citep{Koester:2014}. However, there are several dozen well-characterised polluted WDs with multiple heavy elements in their atmospheres \citep[e.g.][]{Xu:2019, Putirka:2021}, including GD~362 with a total of 19 metals observed at the same time \citep{Becklin:2005, Zuckerman:2007, Xu:2013, Xu:2017, Melis:2017}. With the exception of a few systems rich in water ices \citep[e.g.][]{Farihi:2013, Raddi:2015}, these discoveries have revealed that most WD pollutants are relatively dry and rocky \citep{Jura:2006, Zuckerman:2007, JuraXu:2012, Xu:2019, Swan:2019}, with compositions similar to those of bulk Earth or the CI chondrites ---i.e. the most primordial type of meteorites in the Solar System.

As of today, spectroscopic analyses of polluted WDs have enabled the detection of more than 20 metals, revealing a diverse landscape of extrasolar compositions \citep[e.g.][]{Klein:2021, Xu:2019, Doyle:2019}. Despite these exciting discoveries, the scarcity of polluted WDs with multiple heavy elements in their spectra, combined with the low number of polluted WDs with well-measured elemental abundances, has impeded a statistical study of the pollution phenomenon. In this paper, we address this problem by exploiting the fast and automated Machine Learning (ML) pipeline \cecilia{} (\citealt{BadenasAgusti:2024}, or BA24 thereafter). This pipeline is the first Neural-Network (NN)-based spectral interpolator capable of rapidly estimating the elemental abundances\footnote{In this paper, abundances are expressed as logarithmic number abundance ratios in base 10 relative to Helium, i.e. $\log_{10}$(n(Z)/n(He)).} of intermediate-temperature He-atmosphere polluted WDs (10,000$\leq$\teff$\leq$20,000~K) from optical spectra covering the wavelength range between 3,000~\angs{} and 9,000~\angs{}. More specifically, we use \cecilia{} to measure the atmospheric composition of five polluted WDs with existing \SDSS{} spectra, newly acquired \KeckESI{} observations, and no well-measured abundances in the literature.

This paper is organised as follows. In Section \ref{sec:paper3_motivation}, we motivate the study of He-rich polluted WDs using \cecilia's ML and Bayesian framework. Section \ref{sec:paper3_data} details our target sample and the spectroscopic observations analysed in this work. In Section \ref{sec:paper3_methodology}, we provide an overview of \cecilia{} and  describe our methodology for estimating the atmospheric composition of our targets. Section \ref{sec:paper3_results} presents \cecilia's best-fit models and the geochemical properties of the WD pollutants. In Section \ref{sec:paper3_discussion}, we investigate the limitations of our compositional analysis and discuss potential improvements to our code. 
Finally, we summarise our work and conclude in Section \ref{sec:paper3_conclusions}. 

\section{Motivation} \label{sec:paper3_motivation}

Our choice to study He-atmosphere polluted WDs is driven by two key observational advantages. First, their lower photospheric opacity compared to H-rich WDs makes it easier to observe low levels of metal pollution \citep{Dufour:2012, Klein:2021, Saumon:2022}. Second, their extended convection zones often result in longer metal diffusion timescales, which also facilitates the detection of metal pollution \citep[e.g.][]{Zuckerman:2010}. From a computational perspective, we leverage the fast and automated interpolation capabilities of \cecilia{} to address the limitations of conventional WD characterisation methods. These ``classical'' tools have underpinned the field of WDs for several decades, offering unique insights into the bulk composition of extrasolar material. Nevertheless, they involve time-intensive and manual work, so they would be too impractical and prohibitively expensive to analyse large samples of polluted WDs. For example, in the coming years, multiple wide-field spectroscopic surveys will deliver an unprecedented amount of data to the WD community. This includes the Sloan Digital Sky Survey V (SDSS-V; \citealt{Kollmeier:2017_SDSSV, Chandra:2021}, the Dark Energy Spectroscopic Instrument (DESI; \citealt{DESIa_Cooper2023, DESIa, DESIb}), or WEAVE \citep{Dalton:2014_WEAVE}, which are expected to acquire spectra of about 100,000, 40,000, and 50,000 WDs, respectively, in the near future ---some of which may exhibit signs of metal pollution. This vast amount of data would be intractable with conventional, ``human-in-the-loop'' methods. However, \cecilia{} can obtain preliminary abundance measurements in less than a month using a single GPU \citep{BadenasAgusti:2024}, therefore offering a scalable solution to mine large databases with minimal human supervision.

\begin{table*}
    \centering
    \caption{Main astrophysical properties of the five He-atmosphere polluted WDs studied in this work. References (Ref.): [1] The \Gaia{} Mission \citep{Gaia:2016, Gaia:2023}, [2] The \SDSS{} spectroscopic survey \citep{York:2000, Almeida:2023}, [3] The \Galex{} database \citep{Martin:2005_GALEX}, [4] The \panstarrs{} database \citep{PanSTARRS:2010}, [5] This paper. In particular, \teff{} and \logg{} were obtained from an external fit to \panstarrs{} and \SDSS{} photometry, while the WD mass and the cooling age were derived from the MWDD evolutionary models of \citealt{Bedard:2020} (see Section \ref{sec:paper3_estimation_wd_abundances}), [6] ``MWDD HE'' column in the MWDD (\citealt{Dufour:2016_MWDD}; see Section \ref{sec:paper3_estimation_wd_abundances}), [7] Spectral type from \citealt{Coutu:2019}. \label{tab:paper3_wd_info} }
    \renewcommand{\arraystretch}{1.45}
    \addtolength{\tabcolsep}{-4.55pt}  
    \begin{tabular}{|lc|c|c|c|c|c|}
    \hline
    Property & Ref. & {\bf \wdzerotwothreeone}  & {\bf \wdzeroeightfivenine} &  {\bf \wdoneonezeronine}  & {\bf \wdonethreethreethree}  & {\bf \wdtwothreeoneone}  \\
    \hline \hline
    \multicolumn{7}{|c|}{\it{Other Target Names}}  \\
    \hline
     \Gaia{} DR3 ID         &  [1] & 102350823010868736    & 1037518722660955392   & 3965233688795064832   & 1665473315344805760   & 2650975899537545728\\
    \SDSS{}  ID             &  [2] & J023154.82$+$251259.5 & J085957.20$+$573249.9 & J110957.82$+$131827.9 & J133306.98$+$634936.4 & J231141.58$-$004100.7 \\
    \Galex{}  ID            &  [3] & J023154.8$+$251259    & J085957.2$+$573249    & J110957.8$+$131827   & -                     & J231141.6$-$004100 \\
    \hline 
    \multicolumn{7}{|c|}{\it{Astrometric Properties}}  \\
    \hline 
    R.A. [J2020; h:m:s]        &  [1]  & 02:31:54.82    & 08:59:57.19     & 11:09:57.83     & 13:33:06.99      & 23:11:41.58      \\
    Dec. [J2020; d:m:s]        &  [1]  & +25:12:59.51   & +57:32:50.01    & +13:18:28.08    & +63:49:36.38     & $-$00:41:00.62     \\
    Parallax [mas]             &  [1]  & 3.16$\pm$0.24  & 3.14$\pm$0.27   & 3.35$\pm$0.22   & 3.39$\pm$0.12    & 4.23$\pm$0.18   \\
    Distance [pc]              &  [1]  & 316.59$\pm$24  & 318.94$\pm$27   & 298.36$\pm$20   & 294.94$\pm$11    & 236.46$\pm$10      \\
    $\mu_{{\rm R.A.}}$ [\masy] &  [1]  & 0.41$\pm$0.27  & -4.32$\pm$0.23  & -18.64$\pm$0.27 & -32.54$\pm$0.17  & -18.27$\pm$0.20 \\
    $\mu_{{\rm Dec.}}$ [\masy] &  [1]  & 8.22$\pm$0.27  & -17.99$\pm$0.21 & -32.79$\pm$0.21 & -1.36$\pm$0.13   & -23.86$\pm$0.19 \\
    \hline 
    \multicolumn{7}{|c|}{\it{Photometric Properties}}  \\
    \hline
    \Gaia{} G$_{\rm mag}$      & [1]   & 18.89$\pm$0.04   & 19.11$\pm$0.03    & 18.73$\pm$0.03    & 18.52$\pm$0.01     & 18.71$\pm$0.01   \\
    \SDSS{}  g$_{\rm mag}$     & [2]   & 18.84$\pm$0.01   & 18.97$\pm$0.01    & 18.59$\pm$0.01    & 18.39$\pm$0.01     & 18.62$\pm$0.01   \\
    \panstarrs{} g$_{\rm mag}$ & [4]   & 18.840$\pm$0.004 & 19.03$\pm$0.01    & 18.63$\pm$0.01    & 18.44$\pm$0.01     & 18.650$\pm$0.004 \\
    \hline 
    \multicolumn{7}{|c|}{\it{Physical Properties}}  \\
    \hline
    \teff{} [K]                & [5]  & 12620$\pm$503            & 12677$\pm$722          & 15112$\pm$1688         & 14762$\pm$1340         & 12023$\pm$544 \\
    \logg{} [cgs]              & [5]  & 7.76$_{-0.13}^{+0.14}$   & 7.95$_{-0.15}^{+0.16}$ & 8.09$_{-0.19}^{+0.20}$ & 7.95$_{-0.13}^{+0.14}$ & 8.07$_{-0.09}^{+0.10}$ \\
    Mass [\Msun]               & [5]  & 0.45$_{-0.06}^{+0.07}$   & 0.56$_{-0.08}^{+0.10}$ & 0.64$_{-0.11}^{+0.13}$ & 0.56$_{-0.07}^{+0.08}$ & 0.62$_{-0.06}^{+0.06}$ \\
    Cooling Age [Gyr]          & [5]  & 0.24$_{-0.04}^{+0.06}$   & 0.30$_{-0.07}^{+0.10}$ & 0.23$_{-0.09}^{+0.13}$ & 0.19$_{-0.06}^{+0.08}$ & 0.42$_{-0.07}^{+0.09}$ \\
    Luminosity [$L_\odot$] & [5] & 0.005$_{-0.002}^{+0.002}$ & 0.004$_{-0.002}^{+0.002}$ & 0.007$_{-0.004}^{+0.005}$& 0.007$_{-0.004}^{+0.004}$ & 0.003$_{-0.001}^{+0.001}$ \\
    
    Spectral Type$^\textit{a}$ &      & DB [6]                   & DBZ [7]                & DBAH [6]               & DBZA [7]               &  DBZ [7] \\
    \hline 
    \multicolumn{7}{|c|}{\it{SDSS Observational Log}}  \\
    \hline
    Date [MJD]                 & [2]  & 2005 Nov 04 & 2003 Jan 23 & 2003 Dec 25  & 2000 Apr 05  & 2003 Nov 20 \\
    Spectrograph               & [2]  & \SDSS{}     & \SDSS{}     & \SDSS{}      &   BOSS       & \SDSS{} \\
    Plate ID                   & [2]  & 2399        & 483         & 1751         &   6822       & 381 \\
    MJD ID                     & [2]  & 53764       & 51924       & 53377        &   56711      & 51811 \\
    Fiber ID                   & [2]  & 157         & 463         & 9            &   196        & 103 \\
    Wavelength Coverage [\angs{}] & [2] & 3802-9191   & 3809-9215   & 3810-9206    &  3565-10341  & 3794-9198 \\
    Mean S/N                   & [5]  & 17.03 & 8.60 & 10.45 & 27.85 & 10.91 \\ %at 5,500~\angs{}
    \hline
    \multicolumn{7}{|c|}{\it{Keck/ESI Observational Log}}  \\
    \hline
    Date [UT]                     & [5]  & 2016 Nov 19  & 2016 Nov 18, 2017 Mar 7 & 2017 Mar 7    & 2016 Mar 28       & 2016 Jun 8, 2016 Nov 19 \\
    Time [s]                      & [5]  & 4,800        & 21,000                  & 6,000         & 3,000             & 9,000 \\
    Standard Star                 & [5]  & BD+28\degree4211   & G191-B2B, Feige 34      & HZ 44         & BD+28\degree4211  & BD+28\degree4211 \\
    Wavelength Coverage [\angs{}] & [5]  & 3915-9299    & 3915-9299               & 3915-9299     & 3915-9299         & 3915-9299 \\
    Mean S/N                      & [5]  & 28           & 65                      & 47            & 34                & 45 \\ %at 5,500~\angs{}
    \hline
    \end{tabular}
    \vspace{0.1cm}
    \begin{quote}
         \hspace{-0.17cm}\footnotesize{[\it{a}}]: \footnotesize{The estimated mass for \wdzerotwothreeone{} is below 0.5\Msun, which could point to the binary nature of this system \citep{RebassaMansergas:2011}. This scenario was also proposed by the ML study of \citet{Vincent:2023}, where \wdzerotwothreeone{} was autonomously classified as a DB$+$M dwarf binary based on its \SDSS{} spectrum. Although we cannot fully rule out this hypothesis, we see no visual evidence of it in our spectra (e.g. no large RV variations, or  M-dwarf-like flux modulations in the red part of the spectrum). Moreover, the Renormalized Unit Weight Error (\texttt{ruwe}) of this object is $1.05$, which falls below the \Gaia{} DR3 threshold for unresolved binaries (\texttt{ruwe}$>$1.25; \citealt{Penoyre:2020}).} 
    \end{quote}
\end{table*}

\section{Data} \label{sec:paper3_data}

\subsection{Target Selection}  \label{sec:paper3_wd_selection}

The five He-atmosphere polluted WDs considered in this work are \wdzerotwothreeone, \wdzeroeightfivenine, \wdoneonezeronine, \wdonethreethreethree,  and \wdtwothreeoneone{}. In Table \ref{tab:paper3_wd_info}, we present their full \Gaia{} IDs and summarise their main astrophysical properties and spectroscopic observations.

Our targets were selected from the sample of \citet{KoesterKepler:2015}, who originally identified them as He-rich polluted WDs in the Sloan Digital Sky Survey (\SDSS; \citealt{York:2000}). From this sample, we chose about 20 objects for follow-up at higher resolution and signal-to-noise ratio (S/N) using the Echellette Spectrograph and Imager (ESI) on the Keck II Telescope \citep{Bigelow:1998, Sheinis:2002} (programs:  U067E, U131E, U153E, and U059; PI: B. Zuckerman). Among the observed systems,  our five WDs exhibited the most interesting spectral features and were thus ideal candidates for a detailed abundance analysis with \cecilia. Beyond their polluted nature, we also selected our targets based on two additional criteria. First, their effective temperatures (\teff) and surface gravities (\logg) satisfied \cecilia's allowed parameter bounds (see Table 1 in BA24). Second, they all lacked well-measured photospheric compositions, with only estimates of their H and Ca abundances available in the Montreal White Dwarf Database (MWDD; \citealt{Dufour:2016_MWDD}) from the study of \citet{KoesterKepler:2015}.\footnote{The MWDD (\url{https://www.montrealwhitedwarfdatabase.org/}) is the largest database of spectroscopically confirmed WDs to date.} 

In our final target sample, only \wdoneonezeronine{} had an estimated Ca abundance from \citet{KoesterKepler:2015} ($-$6.46$\pm$0.5) slightly above \cecilia's maximum bound for this metal ($-$7.00). However, we chose to include this object in our analysis because \cecilia{} revises its Ca abundance to be well within its acceptable range (see Table \ref{tab:paper3_mcmc_results}). Morever, \cecilia{} produces accurate fits to the observed Ca spectral lines (\autoref{fig:paper3_fit_wd1109}) and yields a fully converged MCMC posterior distrubution that is not truncated at the upper limit  (\autoref{fig:paper3_corner_wd1109}).\footnote{In line with the case of \wdoneonezeronine{}, systems that would \textit{a priori} be discarded for having \teff{}, \logg, or abundance estimates near or beyond \cecilia's bounds can be retained for atmospheric analysis if their MCMC posterior distributions do not pile up at the limits, but instead remain well inside \cecilia's boundaries.} For the remaining WDs, their published astrophysical properties, including their H and Ca abundances, satisfied \cecilia's allowed parameter space at the time of this writing. 

\subsection{Spectroscopic Observations} 

\subsubsection{\SDSS} 

All five WDs were observed with the \SDSS{} spectrograph (\wdzerotwothreeone, \wdoneonezeronine, \wdzeroeightfivenine, \wdtwothreeoneone) or the upgraded \textit{BOSS} spectrograph (\wdonethreethreethree) mounted on the Sloan Foundation Telescope at Apache Point Observatory \citep{York:2000, Gunn:2006}. Their spectra (in vacuum) were downloaded as FITS files from the \SDSS{} DR18 online database\footnote{\url{https://skyserver.sdss.org/dr18/}} and have a variable resolving power ($R\equiv\lambda/\Delta\lambda$) of R$\approx$1,500 at 3,800~\angs{} and R$\approx$2,500 at 9,000~\angs{}. In our optimisation routine, we assumed a fixed R$=$2,000 for each SDSS dataset (see Section \ref{sec:paper3_estimation_wd_abundances}). 

\begin{figure*}
      \centering
      \includegraphics[width=0.81\linewidth]{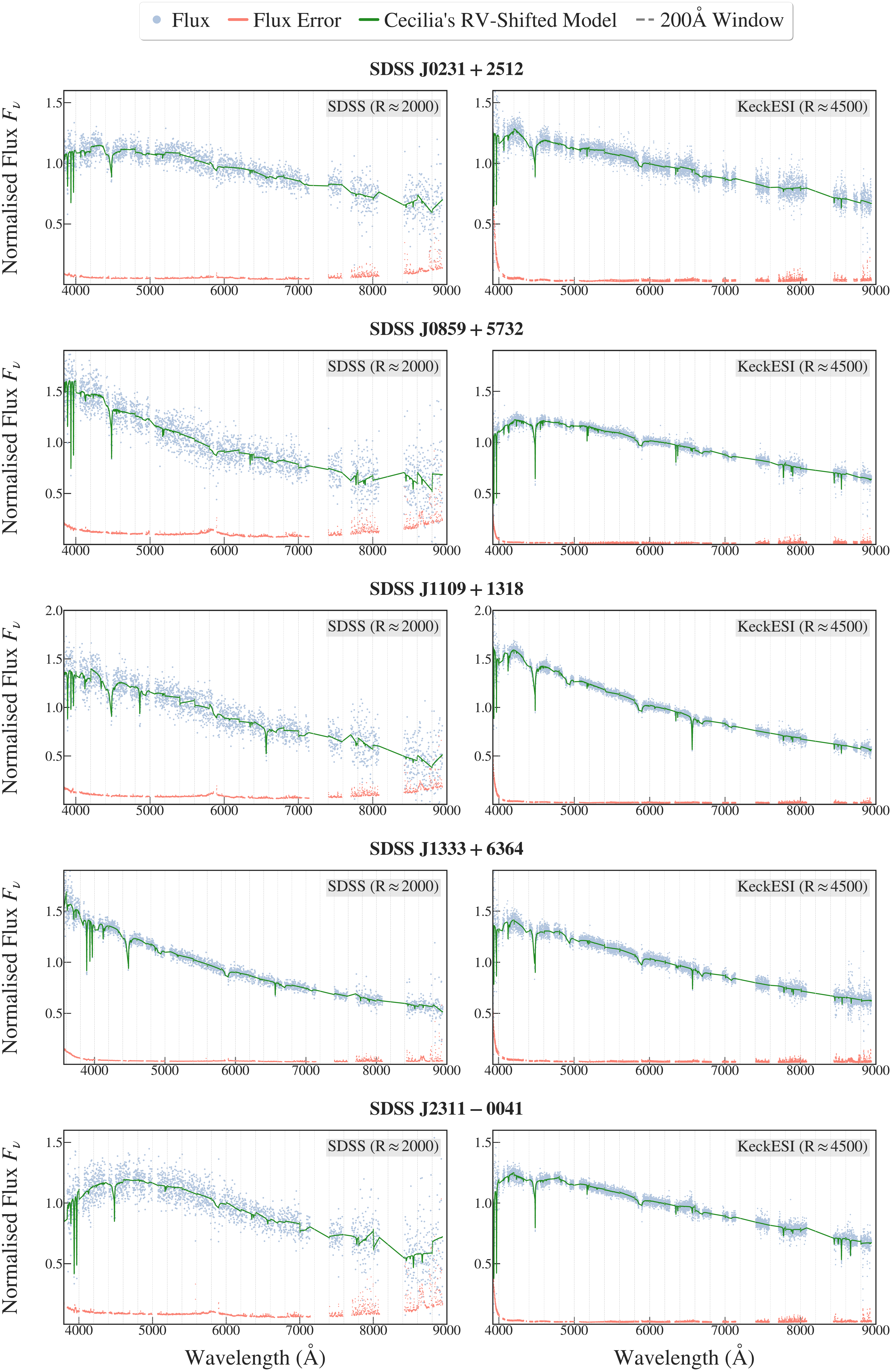}
      \caption{Median-normalised optical spectra of the five polluted WDs in our sample (\textit{left}: \SDSS{} with an assumed fixed resolving power of R$=$2,000 (in vacuum); 
      \textit{right}: \KeckESI{} with an assumed fixed R$=$4,500 in air). The stellar fluxes and their corresponding uncertainties are shown, respectively, in blue and red, while \cecilia's best-fit models are presented in dark green. The data gaps correspond to the discarded wavelength regions described in Section \ref{sec:paper3_data_treatment}, including the strongest He I lines. The grey dashed vertical lines denote the 200~\angs{} spectral windows used during \cecilia{}'s training and optimisation routine.}
      \label{fig:paper3_wd_spectra}
\end{figure*}

\subsubsection{\KeckESI} 

The five polluted WDs in our sample were also observed with the \KeckESI{} spectrograph \citep{Sheinis:2002} across 10 overlapping orders covering the wavelength range 3,900$-$11,000~\angs{}. All datasets (in air) were obtained with a slit width of 1~arcseconds (\arcsec), resulting in a resolving power of about R$=$4,500. The observing logs of the \KeckESI{} observations are presented in Table \ref{tab:paper3_wd_info}.

After acquiring the data, we carried out basic reductions using the \texttt{makee} package,\footnote{\url{https://sites.astro.caltech.edu/~tb/makee/}} including bias subtraction, flat fielding, spectrum extraction, wavelength calibration, and barycentric correction. We performed flux calibration on the spectra by computing their corresponding instrumental sensitivity function. To this end, we used \KeckESI{} observations of spectrophotometric standards listed in the database of the European Southern Observatory,\footnote{\url{https://www.eso.org/sci/observing/tools/standards/spectra/stanlis.html}} obtained spatially and temporally close to the science target observations. For each WD, we computed the wavelength-dependent flux response of the instrument using STIS spectra from the CALSPEC STScI database \citep{Bohlin:2014}.\footnote{\url{https://archive.stsci.edu/hlsps/reference-atlases/cdbs/calspec/}} The instrumental response function was then applied to the extracted spectra of the science targets to obtain a flux-calibrated spectrum. We then stacked all the orders together to generate a continuous spectrum by binning the wavelength, flux, and flux uncertainties over an output wavelength grid spanning the full wavelength range, with wavelength spacing defined by the lowest resolution order to conserve flux. For WDs that were observed multiple times (either during the same night or on different days; see Table \ref{tab:paper3_wd_info}), we followed a similar stacking strategy and combined all their existing observations into a final spectrum with an improved S/N.

\subsection{Data Treatment} \label{sec:paper3_data_treatment}

Next, we visually inspected the \SDSS{} and reduced \KeckESI{} spectra to minimise the impact of spurious effects, instrumental noise, and telluric contamination. We first eliminated scattered outliers, skylines, infinite (\textit{NaN}) flux points, and flux error measurements equal to zero. We then cleaned the \SDSS{} and \KeckESI{} spectra from tellurics by clipping the spectral bands of O$_{2}\gamma$ + O$_{4}$ (air wavelengths: 6,270-6,330~\angs), O$_{2}$B (6,860-6,965~\angs{}), H$_{2}$O (7,143-7,398~\angs{}), O$_{2}$A (7,586-7,703~\angs{}), and H$_{2}$O (8,085-8,420~\angs{} and 8,915-9,000~\angs{}) \citep{Buton:2013}. In addition to these processing steps, we also addressed a well-known limitation of atmosphere models for He-rich WDs cooler than \teff$\lesssim$16,000~K (as is the case for the five targets in our sample) ---namely, their inability to accurately reproduce the broadening of He I lines caused by collisions with neutral particles, such as He and H atoms \citep{Bergeron:2011, GenestBeaulieu:2019_feb, Cukanovaite:2020, Saumon:2022}. Given that our pipeline was trained on these atmosphere models, any inaccuracies in the treatment of He I broadening could propagate to \cecilia's predicted elemental abundances (since the amount of metals in the WD affects the stellar electron density and photospheric opacity, and hence the shapes of the spectral lines; \citealt{Hollands:2017}). Therefore, we sought to mitigate this issue with two complimentary strategies: 

\vspace{5pt}
\begin{enumerate}[noitemsep,topsep=1pt,leftmargin=26pt]
    \item \textit{Clipping of prominent He I absorption lines}: First, we removed the strongest He I features from each spectrum. To perform this task, we applied a symmetric clipping window of 20~\angs{} around the center of the lines, focusing on those at air wavelengths 3,819~\angs, 4,026~\angs, 4,387~\angs, 4,713~\angs, 4,921~\angs, 5,015~\angs, 5,047~\angs, 5,875~\angs, 6,678~\angs, 7,065~\angs, and 7,281~\angs{} \citep{NIST_ASD}. For the He I feature at 4,471~\angs{}, we used an asymmetrical window of 20~\angs{} and 13~\angs{} to the left and to the right of the center, respectively, in order to preserve a potential Mg absorption line near 4,481~\angs. The final \SDSS{} and \KeckESI{} observations used in our analysis are shown in \autoref{fig:paper3_wd_spectra}.\footnote{In a few cases, we intentionally retained a small number of weaker He I absorption lines when they were visible in some (but not all) the spectra, especially if they were close to metal lines. This includes the He I absorption feature at about 4,121~\angs{} in Panel \textit{c} of \autoref{fig:paper3_corner_wd0859}, which is visible in some of the \KeckESI{} spectra, and is close to Fe I and O II transitions. For consistency, we did not remove these weaker He I features when present, but we verified that their inclusion had a negligible effect on \cecilia's final results, including the best-fit elemental abundances.}

    \vspace{5pt}
    \item \textit{Fixed photometric values for \teff{} and \logg{}}: Second, we froze the  effective temperature and surface gravity of each WD to their photometric solution during \cecilia's fitting procedure (see Section \ref{sec:paper3_estimation_wd_abundances}). This decision was not only motivated by the unreliable nature of He-rich WD models below \teff$\lesssim$16,000~K, but also by the strong effect of He I lines in driving \cecilia's spectroscopic results for \teff{} and \logg{}. In particular, although we removed the strongest He I features from our spectra, our clipping method did not eliminate the full extent of their wings, allowing residual He I to affect \cecilia's predictions. In addition to this leftover He I absorption, we also observed that prominent metal lines (e.g. Ca II at about 3,934~\angs{} and 3,969~\angs) could similarly influence the inferred values for \teff{} and \logg{}. However, unlike the He I lines, these metal features could not be removed without losing important information about the WD's elemental abundances. Therefore, given that both residual He I absorption and strong metal lines could bias \cecilia's spectroscopic solution for \teff{} and \logg{}, we adopted a more conservative approach by fixing these two parameters to their photometric estimates. We note that we also explored leaving \teff{} and \logg{} free, but even with informative photometric priors based on external fits to the stellar photometry, \cecilia's results would still deviate significantly from their photometric solution. This behaviour further reinforced our decision to rely on the fixed photometric values for \teff{} and \logg{} in our optimisation routine.
\end{enumerate}
\vspace{5pt}

\section{Methodology} \label{sec:paper3_methodology}

In this Section, we describe our procedure for estimating the elemental abundances of the five polluted WDs. First, we provide a brief summary of \cecilia{} and present several new functionalities  recently added to our code. We then discuss \cecilia's optimisation procedure.

\subsection{An Overview of \cecilia} \label{sec:paper3_cecilia_overview}

In recent years, ML has improved our ability to address complex scientific problems, while also helping to reduce our dependence on time-intensive and manual data analysis techniques. These advancements have impacted all branches of astropyhsics, offering new ways to exploit large astronomical datasets and learn about the fundamental properties of planets and stars. Despite this progress, however, the complex spectral features of polluted WDs have not previously been modeled with ML-based techniques. Therefore, we chose to develop \cecilia{}, the first ML system capable of measuring the main astrophysical properties of He-rich polluted WDs from their spectra. Implemented with the open-source ML package  \texttt{tensorflow}\footnote{\url{https://www.tensorflow.org/?hl=es}} \citep{tensorflow2015-whitepaper, tensorflow:2016}, and trained with MIT's Satori Supercomputer and one NVidia V100 GPU,\footnote{\url{https://mit-satori.github.io/}.} \cecilia's ML architecture is composed of three unsupervised, multi-layered (i.e. ``deep'') Neural Networks (NN): an Autoencoder, a Fully Connected Neural Network (FCNN1), and a Fine-Tuning Fully Connected Neural Network (FT FCNN2).\footnote{The Autoencoder has two components: an Encoder and a Decoder. Together, they are used for data compression and dimensionality reduction. The FCNNs are designed to produce a high-resolution synthetic spectrum from 13 stellar labels (i.e. \teff, \logg, \logHHe, and 10 metal abundances relative to He).} These networks use thousands of computational units (or ``neurons'') and non-linear ``activation'' functions to transform a set of input features (WD properties) into  useful output parameters (a polluted WD spectrum). 

To achieve a good balance between predictive accuracy and computational efficiency, we trained \cecilia{} sequentially using 29 windows of 200~\angs{} in the wavelength range between 3,000~\angs{} and 9,000~\angs. Our training, validation, and testing sets involved more than 22,000 randomly generated synthetic WD properties (or ``labels''), together with their corresponding synthetic spectra (in air wavelengths). As described in BA24, \cecilia's labels consist of 13 independently varied parameters: the effective temperature \teff{} and surface gravity \logg{} of the WD; its logarithmic hydrogen abundance relative to He, \logHHe{}; and 10 additional number abundances for Ca, Mg, Fe, O, Si, Ti, Be, Cr, Mn, and Ni. We also incorporated 14 additional elements (C, N, Li, Na, Al, P, S, Cl, Ar, K, Sc, V, Co, Cu) with their calcium abundance ratio (Z/Ca) scaled to their CI chondritic abundance from \citet{Lodders:2003}. In addition to our choice of 13 stellar labels, we trained \cecilia{} with high-resolution (R$\approx$50,000) synthetic spectra featuring 55,000 points between 3,000~\angs{} and 9,000~\angs{}. These models were generated with the atmosphere code of \citealt{Dufour:2007,Blouin:2018a,Blouin:2018b}, which is a local thermodynamic equilibrium code that self-consistently considers H, He, and metallic species in its equation of state and in the calculation of opacities. Metal lines are included using the Vienna Atomic Line Database (VALD; \citealt{Piskunov:1995,Kupka:1999,Ryabchikova:2015}).    

\begin{figure*}
    \centering
       \includegraphics[width=1\linewidth]{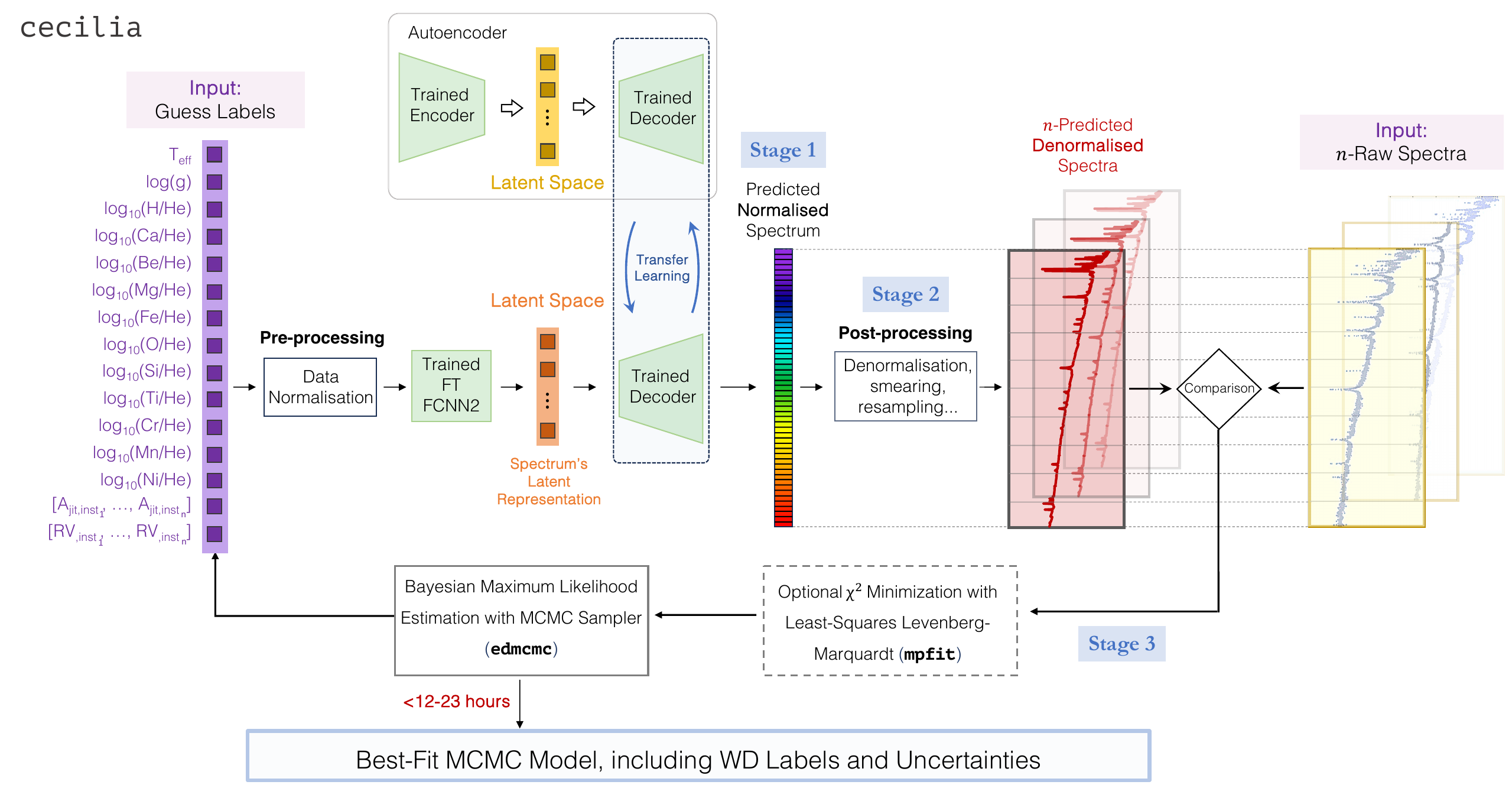}
      \caption{A summary of \cecilia's methodology for estimating the main astrophysical properties (or labels) of polluted He-rich WDs from multiple spectroscopic observations. We refer the reader to \citet{BadenasAgusti:2024} for a more comprehensive description of the pipeline.}
      \label{fig:paper3_cecilia_summary}
\end{figure*}

Once trained, \cecilia{} exploits the speed of NN-based interpolation to rapidly produce high-resolution model spectra in a fully autonomous manner. For example, in the context of this work, \cecilia{} can predict a synthetic \SDSS{} or \KeckESI{} spectrum in 0.17-0.19 seconds, with individual wavelength windows generated in less than 6 miliseconds on average. When factoring in additional computational overheads (e.g. estimation of RV shifts), the total processing time per spectrum reaches 0.35-0.37 seconds, including the evaluation of the MCMC log-likelihood function. This efficiency makes \cecilia{} several orders of magnitude faster than conventional WD atmosphere codes, which require one to three hours on a single CPU core to produce an equivalent synthetic spectrum. As \cecilia{} continues to evolve, further refinements in its ML architecture could help improve its speed, making it more efficient for large-scale WD spectral analysis. We also note that \cecilia's interpolation time is largely unaffected by variations in the input stellar labels (i.e. \teff, \logg, and elemental abundances). This stability is expected, as \cecilia{}'s architecture leverages the learned latent-space representations of its training dataset, therefore bypassing the need for computationally intensive radiative transfer calculations. A summary of \cecilia{}'s framework is presented below and in \autoref{fig:paper3_cecilia_summary}. We refer the reader to BA24 for a more detailed discussion of the pipeline.

\begin{enumerate}
[noitemsep,topsep=1pt,leftmargin=14pt,listparindent=1.5em]
    \vspace{5pt}
    \item \textit{Stage 1}: First, \cecilia{} invokes its trained FT FCNN2 network\footnote{We note that \cecilia{} only employs its FT FCNN2 architecture to generate a spectral prediction. The Autoencoder and FCNN1 are only used during the training stage to improve the overall learning accuracy of our pipeline.} to produce a preliminary, high-resolution synthetic spectrum based on user-defined initial guesses for the 13 stellar labels of the WD (i.e. \teff, \logg, and 11 logarithmic elemental abundances for H, Ca, Mg, Fe, O, Si, Ti, Be, Cr, Mn, and Ni). If any of these labels are unknown (as would likely be the case for a newly discovered polluted WD), \cecilia{} assumes chondritic abundance ratios except for H and Ca, for which it adopts their mean abundance ratio from its training set as initial guesses. We note that the assumption of chondritic abundance ratios is often used to approximate the photospheric abundances of elements when, for various reasons, these cannot be derived directly from a WD spectrum \citep{Dufour:2007, Coutu:2019}. Although several exceptions have been found ---such as water-rich material  \citep{Farihi:2013, Raddi:2015, GentileFusillo:2017}, or differentiated bodies with crustal and core-like compositions \citep{Zuckerman:2011, Jura:2013, Harrison:2018}---  this assumption is often used given the similarity between the bulk composition of WD pollutants, and those of bulk Earth and the CI chondrites \citep{Allegre:2001, Lodders:2003}, especially for rock-forming elements like Ca, Mg, Fe, Si, and O \citep{Zuckerman:2007, Xu:2019, Doyle:2023}.
    
    \vspace{5pt}
    \item \textit{Stage 2}: Second, \cecilia{} implements several data processing techinques to enable a meaningful comparison between its initial prediction and the observed polluted WD spectrum. For example, it (i) denormalises its raw ML output; (ii) smears and resamples its denormalised prediction to the the resolving power and wavelength grid of the observed spectrum, respectively; (iii) applies a radial velocity shift (or RV thereafter) to its prediction; and (iv) employs a linear model (or alternatively, an $n$-th degree polynomial as decided by the user) to correct the spectral ``jumps'' arising from the training of \cecilia{} in independent windows of 200~\angs{}. 

    \vspace{5pt}
    \item \textit{Stage 3}: Finally, \cecilia{} employs a combination of frequentist and Bayesian statistical techniques to optimise its smeared, downsampled, and slope-corrected ML prediction. In particular, it incorporates two fitting routines: an optional and fast non-linear least-squares Levenberg-Marquardt method implemented by the Python \texttt{mpfit}\footnote{\url{https://github.com/segasai/astrolibpy/blob/master/mpfit/mpfit.py}} library \citep{More:1978, Markwardt:2009}, and a  differential evolution Markov Chain Monte Carlo (MCMC) sampler executed by the Python \texttt{edmcmc} package \citep{Vanderburg:2021_edmcmc, Ter:2006}. If the user runs \texttt{mpfit}, the resulting best-fit parameters are fed into the MCMC as improved initial guesses. 
\end{enumerate}
\vspace{5pt}

For a given white dwarf, \cecilia{} takes, on average, less than a day to produce a complete spectroscopic solution using a single GPU. Moreover, \cecilia{} can be scaled to simultaneously process multiple objects if additional GPUs are available, making it a valuable tool in the era of large-scale astronomical surveys. The total execution time is primarily dependent on the computational cost associated to the ML-based spectral interpolation process, rather than on the spectral resolution or S/N of the observations. As discussed in BA24, \cecilia{}'s  retrieval accuracy  can be  $\lesssim$0.1~dex for up to 10 chemical elements (H, Ca, Mg, Fe, O, Si, Ti, Be, Cr, Mn, and Ni), with Be appearing the hardest element to constrain.\footnote{\cecilia's performance is discussed in more detail in Section 5 of BA24, based on an analysis of synthetic and real spectroscopic observations.} This performance is comparable to that of conventional techniques, which usually yield uncertainties of about 0.10~dex-0.20~dex for He-dominated polluted WDs, both in the UV and in the optical, as well as with different types of WD atmosphere models \citep[e.g.][]{Doyle:2023, Klein:2021, Izquierdo:2020, Raddi:2015, Wilson:2015, Jura:2012, Zuckerman:2007}.  

Beyond \cecilia's predictive capabilities, our code differs from conventional WD analysis techniques in two important ways. First, it can quickly fit the spectra of polluted WDs without the need for human supervision. This sets it apart from classical methods, which can be slow and time-intensive due to their reliance on manual and iterative fitting procedures. Second, \cecilia{} provides a Bayesian treatment of observational evidence and parameter uncertainties. Through its MCMC, it maximises the log-likelihood function of the free model parameters ($\ln(\mathcal{L})$), while also incorporating the information encapsulated in the spectrum and any ``prior'' assumptions about the properties of the star. This approach allows for a thorough numerical exploration of the multidimensional parameter space of the stellar labels, yielding full posterior probability distributions, robust statistical errors, and insights into potential correlations between model parameters. These degeneracies are often visualised in scatterplot matrices, or ``corner plots,'' which provide the two-dimensional marginal posterior of a pair of labels alongside their respective one-dimensional histogram distributions (e.g. see \autoref{fig:paper3_corner_wd0859}). Finally, \cecilia{} fits the \textit{entire} spectrum simultaneously to determine the 13 stellar labels of the WD. This differs from some classical methods designed to calculate average abundance measurements based on individual, line-by-line fits to visible absorption lines.\footnote{We note that other conventional (i.e. non-AI-based) methods are also capable of fitting the entire spectrum simultaneously. For instance, \citealt{Bhattacharjee:2025} recently employed a least-squares minimisation approach to model the spectra of DAZ and DBZ white dwarfs.}

\subsection{\cecilia's Improved Capabilities} \label{sec:paper3_cecilia_upgrades}

In this paper, we present an improved version of \cecilia's optimisation procedure (see \autoref{fig:paper3_cecilia_summary}). To begin with, we have modified the MCMC to sample the elemental abundances in linear space. This is different to the pipeline presented in BA24, where logarithmic abundances were used to map out the likelihood distribution of the model parameters. To implement this change, we have updated \cecilia{} to: (i) adjust the allowed parameter bounds to linear values (see Table 1 in BA24); (ii) convert the user's initial abundance guesses from log to linear space; and (iii) impose a lower bound of zero on all abundances (representing the complete absence of an element in the atmosphere of the WD). A practical motivation for this switch is that sampling in linear space allows the MCMC to explore very low abundance values and \textit{still} achieve convergence. This is different from our previous log-abundance approach, which could cause the MCMC to diverge towards negative infinity in the case of weak or non-detections. Another advantage of our new approach is that it opens the door to computing statistically robust upper limits on the abundances of the undetected elements. However, we do not attempt to calculate these limits in this work due to computational limitations related to the sparseness of \cecilia's training set at very low abundance values.\footnote{We note that our new parametrisation modifies the implicit prior used in our sampling. Indeed, sampling uniformly in logarithmic space (as done in BA24) corresponds to an implicit prior proportional to $1/x$ in linear space, which favours smaller abundance values. In contrast, our new approach samples uniformly in linear space, which corresponds to an implicit prior that increases exponentially with $\log x$ and favours higher abundances. This distinction is only important when the data provide weak or no constraints on the elemental abundances. Indeed, when the observations are sufficiently informative, the posterior is dominated by the likelihood and the influence of the prior becomes negligible.}

In addition to sampling the elemental abundances in linear space, we have also implemented two important changes to \cecilia{}'s log-likelihood function. First, we have modified our code to allow for a simultaneous fitting of $N_{\rm spec}$ spectra from multiple instrumental facilities ---regardless of their intrinsic characteristics (e.g. resolving power, S/N). Second, we have introduced a jitter term $A_{\rm jit}^{s}$ (where $s$ denotes the instrument that acquired the spectrum) to account for unknown sources of noise in the data and in the atmosphere models used to train \cecilia. The jitter parameter employs the observed scatter in the spectrum to estimate what the uncertainties of the stellar flux should be, lowering or increasing them accordingly if necessary \citep{Ford:2006}. Therefore, it regulates the contribution of each spectrum to the final solution and allows \cecilia{} to robustly fit multiple spectra simultaneously, even if some datasets are significantly poorer than the rest. Ideally, the jitter term should be close to 1 when the flux errors are well estimated prior to initiating \cecilia's optimisation routine. As an additional diagnostic metric, our code also computes the reduced chi-squared ($\chi_{\rm reduced}^{2}$) for the best-fit solution associated to each spectrum, with the expectation that $\chi_{\rm reduced}^{2}\approx1$ when all the observations have contributed similarly to the best-fit solution.

With all the aforementioned changes, \cecilia's new log-likelihood MCMC function is given by Eq. \ref{eq:paper3_mcmc_loglike}, where $f^{s,i}_{\rm synth, corr}$ denotes \cecilia{}'s  prediction for pixel $i$ from instrument $s$ (after denormalisation, resampling, resolution downgrading, RV-shifting, and slope correction). The terms $f^{s,i}_{\rm obs}$ and $f^{s,i}_{\rm obs, err}$  represent the observed stellar flux and its corresponding error, while the priors $\pi_{0, {\rm phot}}(T_{\rm eff}, \log g)$ and $\pi_{0, {\rm chondr}}(X_k)$ are optional contributions to \cecilia{}'s spectral model, applied to the parameters \teff{} and \logg{}, and to the elemental abundances $X_k$, respectively. The photometric priors may be used when $T_{\rm eff}$ and $\log g$ are treated as free model parameters and have reliable external constraints (e.g., from existing photometry).\footnote{The current implementation of \cecilia{} does not predict $T_{\rm eff}$ and $\log g$ from photometric observations. Therefore, our code assumes that the user has reasonably good constraints on these two parameters from external photometric fits.} The chondritic priors, in turn, are designed to limit \cecilia's exploration of the parameter space and help with MCMC chain convergence.

\begin{equation}
    \begin{aligned}
    \ln \mathscr{L} = -\frac{1}{2} \Bigg\{ 
    & \sum_{s=1}^{N_{\rm spec}} \sum_{i=1}^{N_{\rm points}^{(s)}} \left[ 
    \left( \frac{f^{s,i}_{\rm obs} - f^{s,i}_{\rm synth, corr}}{A^{s}_{\rm jit} \cdot f^{s,i}_{\rm obs, err}} \right)^2 
    + 2 \ln \left( A^{s}_{\rm jit} \cdot f^{s,i}_{\rm obs, err} \right)
    \right] \\
    & + \sum_{j=1}^{N_{\rm phot}} \pi_{0, {\rm phot}}(T_{\rm eff}, \log g) 
    + \sum_{k=1}^{N_{\rm elem}} \pi_{0, {\rm chondr}}(X_k) 
    \Bigg\}
    \label{eq:paper3_mcmc_loglike}
    \end{aligned}
\end{equation}

In general, the influence of a chondritic prior on an elemental abundance is determined by its inverse variance, $1/\sigma_{\rm prior, chondr}^2$, where $\sigma_{\rm prior, chondr}$ denotes the standard deviation (or width) of the prior distribution $\pi_{0, {\rm chondr}}$. Assuming that both the prior and the likelihood are Gaussian, the posterior uncertainty, $\sigma_{\rm post}$, is given by

\begin{equation}
    \begin{aligned}
    \frac{1}{\sigma_{\rm post}^2} = \frac{1}{\sigma_{\rm obs}^2} + \frac{1}{\sigma_{\rm prior, chondr}^2},
    \label{eq:paper3_posterior_uncertainty}
    \end{aligned}
\end{equation}
where $\sigma_{\rm obs}$ represents the observational uncertainty. For elements that are typically easy to detect (e.g., Ca, Mg, Si), we recommend adopting broad chondritic priors (i.e., a large $\sigma_{\rm prior, chondr}$) so that $1/\sigma_{\rm prior, chondr}^2$ is very small compared to $1/\sigma_{\rm obs}^2$ and $\sigma_{\rm post}\approx\sigma_{\rm obs}$. This approach minimises the effect of the prior and ensures that the posterior uncertainty is dominated by the observations. In contrast, for elements that are hard to detect, we suggest using narrower chondritic priors (i.e., a small $\sigma_{\rm prior, chondr}$) to facilitate MCMC convergence; in these cases, if an element is undetected, we recommend not reporting its predicted \cecilia{} abundance. In this work, we discuss our choice of prior widths in Section \ref{sec:paper3_estimation_wd_abundances} and Table \ref{tab:paper3_priors}.

\subsection{Estimation of White Dwarf Elemental Abundances}  \label{sec:paper3_estimation_wd_abundances}

We measured the elemental abundances of the five polluted WDs by performing a joint MCMC fit to their \KeckESI{} and \SDSS{} spectra. Our \cecilia{} MCMC model consisted of 17 parameters, namely: the 13 stellar labels underlying \cecilia's training set (i.e. \teff, \logg, and the logarithmic abundances of H, Ca, Mg, Fe, O, Si, Ti, Be, Cr, Mn, and Ni relative to He); 2 jitter terms (\AjitSDSS, \AjitKeck); and 2 RV shifts (\RVSDSS, \RVKeck).\footnote{Although we would expect the RV of a WD to remain consistent across its spectra, we considered separate RV shifts for each dataset to account for known systematic offsets in our calibrations arising from different observational set-ups, instrumental effects, etc. As explained in the text, we do not report or provide a physical interpretation of our best-fit RV results; instead, we only treat them as nuisance parameters in our model.} As described in Section \ref{sec:paper3_data_treatment}, we decided to freeze \teff{} and \logg{} to their photometric solution, and only fitted the remaining 15 parameters.

To prepare our initial guesses, we adopted the \teff, \logg, H, and Ca results of Coutu et al. (2019; C19), whenever possible.\footnote{In C19, the authors studied 1,023 DBZ/DZ(A) WDs with the same atmosphere code used to generate \cecilia's training set; therefore, their results represent the most self-consistent initial guesses for our Bayesian analysis.} There were two systems, however, lacking C19 measurements: \wdzerotwothreeone{} and \wdoneonezeronine. For the former, we took as initial guesses the  \teff{} and \logg{} values derived from a fit to \panstarrs{} photometry (see the ``MWDD He'' column in the MWDD), and the H and Ca results from \citet{KoesterKepler:2015}. For \wdoneonezeronine, we used the \teff{} and \logg{} photometric solution of \citet{GenestBeaulieu:2019_sep}, and the H and Ca abundances of \citet{KoesterKepler:2015}. Then, for the remaining (and unknown) elemental abundances, we assumed chondritic abundances relative to our initial guess for Ca. Finally, we set the jitter terms to 1 (dimensionless) and the RV shifts to 0~km/s.

After building our spectral model, we executed \cecilia's MCMC twice per WD, using the log-likelihood in Eq. \ref{eq:paper3_mcmc_loglike} and a linear function to minimise the spectral jumps between training windows of 200~\angs. Our MCMC hyperparameters consisted of $n_{\rm walkers}$=50~walkers (or chains), $n_{\rm draws}$=3000~draws (number of steps per chain), and a 20$\%$ ``burn-in'' phase aimed at removing non-stationary solutions (i.e. $n_{\rm burn}=0.2n_{\rm draws}$=600~draws).\footnote{The use of multiple walkers (or chains) ensures extensive sampling of the parameter space. Over time, the cumulative distribution of the walkers and draws should approach the final posterior distribution.} Similarly to BA24, we initialised the positions of the walkers using Gaussian balls with standard deviations of 0.01~dex for the elemental abundances, 1~km/s for the two RV shifts, and 0.01 for the jitter terms. We then imposed priors on our free model parameters to limit \cecilia's exploration of the parameter space and facilitate MCMC chain convergence. In particular, we applied uniform priors to the Ca abundance, the RV shifts, and the jitter terms, allowing their values to vary within the bounds defined by \cecilia's training set. For all other elemental abundances, we adopted truncated Gaussian priors informed by the chondritic abundances of \citet{Lodders:2003}. To do this, we set the mean of each distribution to the difference between the element's chondritic abundance and that of calcium. We then applied narrow widths of 0.5~dex for elements that are typically hard to detect (Ti, Be, Cr, Mn, Ni), and broader widths of 2~dex to more commonly observed elements (H, Mg, Fe, O, Si). A summary of our priors is presented in Table \ref{tab:paper3_priors}.

\begin{table}
    \centering
    \caption{Adopted prior distributions for \cecilia's model parameters (see Section \ref{sec:paper3_estimation_wd_abundances}). The symbols $\mathcal{T}$ and $\mathcal{U}$ denote truncated Gaussian and uniform priors, respectively. Where applicable, the mean $\mu$ represents the offset between the chondritic abundance of element $X$ and that of calcium, defined as $\mu = \log(\mathrm{Ca}/\mathrm{He}) - \log(X/\mathrm{He})$ based on the chondritic values of \citet{Lodders:2003}. The standard deviation $\sigma$ specifies the width of the truncated Gaussian distribution. The bounds of the elemental abundances are given in logarithmic base 10. }
    \renewcommand{\arraystretch}{1.3}
    \addtolength{\tabcolsep}{-2pt} 
    \begin{tabular}{|l|ccccc|}
    \hline
    Parameter         & Prior         & Min.  & Max.  & Mean ($\mu$) & Width ($\sigma$) \\
    \hline \hline
    \teff{}    [K]     & \multicolumn{5}{c|}{No Prior Used -- Fixed Parameter} \\
    \logg{}    [cgs]   & \multicolumn{5}{c|}{No Prior Used -- Fixed Parameter} \\
    \logHHe{}  [dex]   & $\mathcal{T}$  & -7.00  & -3.00    & -1.96 & 2   \\
    \logBeHe{} [dex]   & $\mathcal{T}$  & -23.85 & -5.61    & 4.91  & 0.5 \\
    \logOHe{}  [dex]   & $\mathcal{T}$  & -17.46 &  1.25    & -2.10 & 2   \\
    \logMgHe{} [dex]   & $\mathcal{T}$  & -16.89 & -0.17    & -1.24 & 2   \\
    \logSiHe{} [dex]   & $\mathcal{T}$  & -17.42 & -0.51    & -1.22 & 2   \\
    \logCaHe{} [dex]   & $\mathcal{U}$  & -12.00 & -7.00    & --    & --  \\
    \logTiHe{} [dex]   & $\mathcal{T}$  & -19.35 & -2.32    & 1.40  & 0.5 \\
    \logCrHe{} [dex]   & $\mathcal{T}$  & -19.61 & -1.71    & 0.66  & 0.5 \\
    \logMnHe{} [dex]   & $\mathcal{T}$  & -20.08 & -1.65    & 0.82  & 0.5 \\
    \logFeHe{} [dex]   & $\mathcal{T}$  & -18.20 &  0.18    & -1.16 & 2   \\
    \logNiHe{} [dex]   & $\mathcal{T}$  & -18.96 & -1.66    & 0.10  & 0.5 \\
    \RVSDSS{}  [km/s]  & $\mathcal{U}$  & -500   & 500      & --    & --  \\
    \RVKeck{}  [km/s]  & $\mathcal{U}$  & -500   & 500      & --    & --  \\
    \AjitSDSS{} [-]    & $\mathcal{U}$  & 0      & $+\infty$ & --   & --  \\
    \AjitKeck{} [-]    & $\mathcal{U}$  & 0      & $+\infty$ & --   & --  \\
    \hline
    \end{tabular}
    \label{tab:paper3_priors}
\end{table}

For our first MCMC, we fitted all model parameters except \teff{} and \logg{}, which we froze to our initial guesses. We then recycled \cecilia's best-fit results to run a second MCMC with improved values for \teff{} and \logg{} (see Tables \ref{tab:paper3_wd_info}-\ref{tab:paper3_mcmc_results}). To refine these two parameters, we fitted the \SDSS{} and \panstarrs{} photometry of each star following the procedure of C19, but correcting it for interstellar extinction using the \texttt{stilism} reddening models.\footnote{\url{https://stilism.obspm.fr/}. For WDs located beyond d$\gtrapprox$100~pc, it is important to account for interstellar extinction by de-reddening the photometry with 3D dust maps \citep{Coutu:2019, GenestBeaulieu:2019_feb, GentileFusillo:2019}.} Our \cecilia{} MCMC analysis took  between 12.26\,hr (\wdoneonezeronine) and 22.68\,hr (\wdzeroeightfivenine) to complete. We then assessed MCMC chain convergence by computing the Gelman-Rubin potential scale reduction factor $\hat{R}$ for each free model parameter and ensuring that it satisfied $\hat{R}<1.02$ \citep{GelmanRubin:1992, Gelman:2004}.

Upon concluding our fits, we considered an element to be detected if its elemental abundance had a statistical (i.e. MCMC) uncertaintiy lower than than or equal to an assumed detectability threshold of \sigmathresh. We also validated this cut-off empirically by confirming the presence of its absorption feature(s) by eye. Then, for each detected element, we used \cecilia's predicted abundance to estimate its mass ratio relative to He ($R_{\rm ratio, Z}$; Eq. \ref{eq:paper3_metal_mass_ratio}), its mass in the stellar atmosphere  ($M_{\rm Z}$; Eq. \ref{eq:paper3_metal_mass}), and its accretion rate ($\dot{M}_{\rm Z}$; Eq. \ref{eq:paper3_metal_accretion_rate}), together with their $1\sigma$ errors from their 16th, 50th, and 84th quantiles of their MCMC posterior distributions. More specifically, we first calculated the mass ratio of a given metal $Z$ with
 
\begin{equation}
    R_{\rm ratio, Z} = n_{\rm Z} \cdot \frac{u_{\rm Z}}{u_{\rm He}}
    \label{eq:paper3_metal_mass_ratio}.
\end{equation}
where $u_{\rm Z}$ represents the atomic mass of the metal, $u_{\rm He}$ is the atomic mass of He (i.e. $u_{\rm He}=4$ atomic mass units, or amu), and $n_{\rm Z}$ is the metal elemental abundance in linear space (as measured by \cecilia). Second, we determined the mass of a given metal ($M_{\rm Z}$) by invoking the mass of the stellar Convection Zone ($M_{\rm cvz}$). In He-atmosphere WDs, CVZs tend to be very large \citep{Koester:2009} and can therefore be used as a proxy for the total mass of He in the atmosphere, 
\begin{equation}
    M_{\rm Z} = M_{\rm cvz} \cdot R_{\rm ratio, Z}.
    \label{eq:paper3_metal_mass}
\end{equation}

To estimate \Mcvz{}, we employed the evolutionary models of \citet{Bedard:2020} using our improved photometric solution for \teff{} and \logg{}.\footnote{The models of \citet{Bedard:2020} are publicly available in the MWDD (\url{https://www.montrealwhitedwarfdatabase.org/evolution.html}).} We then  inferred the accretion rate of each detected metal ($\dot{M}_{\rm Z}$) by taking the ratio of its total mass ($M_{\rm Z}$) to its gravitational sinking timescale (\tauZ{}), where \tauZ{} was obtained from the public diffusion models of the MWDD,\footnote{Eq. \ref{eq:paper3_metal_accretion_rate} quantifies the rate at which metals diffuse out of the convective layer, and it only corresponds to the true accretion rate under the assumption of steady-state equilibrium, that is, when the influx of metals is balanced by their diffusion in the CVZ. As discussed in Section \ref{sec:paper3_results}, it is likely that our targets are in the build-up or steady-state phase, which makes Eq. \ref{eq:paper3_metal_accretion_rate} a reasonable approximation.} 
\begin{equation}
  \dot{M}_{\rm Z} = \frac{M_{\rm Z}}{\tau_{\rm Z}}.
  \label{eq:paper3_metal_accretion_rate}
\end{equation} 

Finally, for each WD in our sample, we computed a lower limit on the total metal accretion rate ($\sum \dot{M}_{\rm Z}$), along iwth the total mass of metals in the atmosphere ($\sum M_{Z}$). From the latter, we derived the fractional mass contribution of each element ($\rm{MF}_{\rm Z}$) using Eq. \ref{eq:paper3_metal_mass_fraction}. For this calculation, we considered the 10 metals fitted by \cecilia{}, with the best-fit masses of the undetected elements constrained via our chondritic abundance priors. 

\begin{equation}
  \rm{MF}_{\rm Z} = \frac{M_{\rm Z}}{\sum M_{Z}}.
  \label{eq:paper3_metal_mass_fraction}
\end{equation}

\subsection{Estimation of Pollutants' Elemental Abundances}  \label{sec:paper3_estimation_pb_abundances}
 
The photospheric abundances of a WD can be used to probe the bulk composition of its pollutant through a set of well-known WD accretion and diffusion equations \citep{Dupuis:1992, Dupuis:1993a, Dupuis:1993b, Koester:2009}. Assuming a constant accretion rate and the engulfment of a single pollutant,\footnote{Previous work has investigated the accretion of multiple bodies (e.g. small asteroids) with different bulk compositions \citep[e.g.][]{Jura:2008, Wyatt:2014, TurnerWyatt:2019, Trierweiler:2022}.  However, for simplicity, we assume that our WDs accreted a single large object.} we evaluated these equations for three different phases. First, there is a  ``build-up'' (or ``increasing'') phase, when the polluting body has just begun to accrete onto the WD's photosphere. During this stage, the number abundance ratios \nA{} and \nB{} of the pollutant for two different metals $A$ and $B$ are assumed to correspond to the elemental abundances in the atmosphere of the WD, 
\begin{equation}
    \left(\frac{n_{{\rm A}}}{n_{{\rm B}}}\right)_{{\rm P}}=\left(\frac{n_{{\rm A}}}{n_{{\rm B}}}\right)_{{\rm WD}},
    \label{eq:paper3_buildup}
\end{equation}
where the subscript ``WD'' alludes to the observed number abundances in the star ---in our case, obtained from \cecilia{} (see Table \ref{tab:paper3_mcmc_results})--- and ``P'' refers to the inferred composition of the polluting body. After the increasing phase, there is ``steady-state'' phase, {when the effects of accretion and diffusion are (almost) in balance in the convective layer.} During this period, the number abundance ratios of the pollutant are assumed to match those of the star when corrected by their respective diffusion timescales, 
\begin{equation}
    \left(\frac{n_{{\rm A}}}{n_{{\rm B}}}\right)_{{\rm P}}=\left(\frac{n_{{\rm A}}}{n_{{\rm B}}}\cdot\frac{\tau_{{\rm B}}}{\tau_{{\rm A}}}\right)_{{\rm WD}}.
    \label{eq:paper3_steady}
\end{equation}

The last phase of accretion is known as the ``declining'' (or ``decreasing'') phase, when the polluting material is no longer replenished in the stellar photosphere and sinks donwards into the interior, hence disappearing from view. During this phase, lightweight elements (i.e. H and He) remain in the outermost layers of the WD, while heavier elements rapidly sink below the photosphere. If $t$ is the time elapsed since the end of accretion, the ratio of heavy elements in the polluting body can be inferred from the  stellar abundance ratios after introducing an exponential decay term, 
\begin{equation}
    \left(\frac{n_{{\rm A}}}{n_{{\rm B}}}\right)_{{\rm P}}=\left(\frac{n_{{\rm A}}}{n_{{\rm B}}}\cdot\frac{e^{-t/\tau_{{\rm B}}}}{e^{-t/\tau_{{\rm A}}}}\right)_{{\rm WD}}.
    \label{eq:paper3_decay}
\end{equation}

\begin{table*} 
    \centering
    \caption{\cecilia's best-fit parameters for the five polluted WDs in our sample, obtained from a joint MCMC fit to their \SDSS{} and \KeckESI{} spectra. For clarity, we only report the abundances of the detected elements, as \cecilia's results for the remaining chemical species are dominated by our choice of chondritic priors. For each detection, we provide three sources of uncertainty: the statistical errors from \cecilia's MCMC (\sigmastat; left parenthesis), the systematic errors caused by imperfections in \cecilia's ML interpolation and our training models (\sigmasys; center), and the total uncertainties (\sigmatot; right), with the latter computed as the quadrature sum of \sigmastat{} and \sigmasys. In all cases, \sigmatot{} is consistently below our assumed noise floor (\sigmafloor), so we follow the methodology of BA24 and enforce \sigmatot$=\sigma_{\text{floor}}$. In this Table, we denote \cecilia's fixed model parameters with a dagger ($\dagger$). We also exclude our results for the \RVSDSS{} and \RVKeck{} terms, as  \cecilia{} does not account for the stellar barycentric motion and gravitational redshift during its fitting procedure. }
    \label{tab:paper3_mcmc_results}
    \renewcommand{\arraystretch}{2}
    \addtolength{\tabcolsep}{-4pt} 
    \begin{tabular}{|l|c|c|c|c|c|}
    \hline
    Parameter & {\bf \wdzerotwothreeone} & {\bf \wdzeroeightfivenine} & {\bf \wdoneonezeronine} & {\bf \wdonethreethreethree} & {\bf \wdtwothreeoneone} \\
    \hline \hline
    \teff{}$^\dagger$ [K] & 12620$\pm$503 & 12677$\pm$722 & 15112$\pm$1688 & 14762$\pm$1340 & 12023$\pm$544 \\
    \logg{}$^\dagger$ [cgs] & 7.76$^{+0.14}_{-0.13}$ & 7.95$^{+0.16}_{-0.15}$ & 8.09$^{+0.20}_{-0.19}$ & 7.95$^{+0.14}_{-0.13}$ & 8.07$^{+0.10}_{-0.09}$ \\
    \logHHe{} [dex]  & -- & -6.60$^{+0.08}_{-0.09}${\footnotesize{($\pm 0.04$)}}{\footnotesize{($\pm 0.20$)}} & -4.33$^{+0.03}_{-0.04}${\footnotesize{($\pm 0.02$)}}{\footnotesize{($\pm 0.20$)}} & -5.49$^{+0.05}_{-0.05}${\footnotesize{($\pm 0.03$)}}{\footnotesize{($\pm 0.20$)}} & -6.26$^{+0.07}_{-0.09}${\footnotesize{($\pm 0.04$)}}{\footnotesize{($\pm 0.20$)}} \\
    \logOHe{} [dex]  & -- & -5.35$^{+0.03}_{-0.03}${\footnotesize{($\pm 0.12$)}}{\footnotesize{($\pm 0.20$)}} & -5.66$^{+0.06}_{-0.06}${\footnotesize{($\pm 0.13$)}}{\footnotesize{($\pm 0.20$)}} & -5.85$^{+0.08}_{-0.08}${\footnotesize{($\pm 0.13$)}}{\footnotesize{($\pm 0.20$)}} & -5.64$^{+0.04}_{-0.04}${\footnotesize{($\pm 0.13$)}}{\footnotesize{($\pm 0.20$)}} \\
    \logMgHe{} [dex] & --7.22$^{+0.06}_{-0.07}${\footnotesize{($\pm 0.08$)}}{\footnotesize{($\pm 0.20$)}} & -6.24$^{+0.02}_{-0.03}${\footnotesize{($\pm 0.07$)}}{\footnotesize{($\pm 0.20$)}} & -6.46$^{+0.05}_{-0.05}${\footnotesize{($\pm 0.07$)}}{\footnotesize{($\pm 0.20$)}} & -6.54$^{+0.07}_{-0.09}${\footnotesize{($\pm 0.07$)}}{\footnotesize{($\pm 0.20$)}} & -6.57$^{+0.02}_{-0.02}${\footnotesize{($\pm 0.07$)}}{\footnotesize{($\pm 0.20$)}} \\
    \logSiHe{} [dex] & -- & -6.48$^{+0.03}_{-0.03}${\footnotesize{($\pm 0.08$)}}{\footnotesize{($\pm 0.20$)}} & -6.72$^{+0.06}_{-0.06}${\footnotesize{($\pm 0.09$)}}{\footnotesize{($\pm 0.20$)}} & -- & -6.78$^{+0.07}_{-0.08}${\footnotesize{($\pm 0.09$)}}{\footnotesize{($\pm 0.20$)}} \\
    \logCaHe{} [dex] & -8.58$^{+0.08}_{-0.08}${\footnotesize{($\pm 0.09$)}}{\footnotesize{($\pm 0.20$)}} & -7.86$^{+0.04}_{-0.04}${\footnotesize{($\pm 0.09$)}}{\footnotesize{($\pm 0.20$)}} & -7.34$^{+0.07}_{-0.08}${\footnotesize{($\pm 0.09$)}}{\footnotesize{($\pm 0.20$)}} & -7.78$^{+0.06}_{-0.06}${\footnotesize{($\pm 0.09$)}}{\footnotesize{($\pm 0.20$)}} & -7.94$^{+0.03}_{-0.03}${\footnotesize{($\pm 0.09$)}}{\footnotesize{($\pm 0.20$)}} \\
    \logFeHe{} [dex] & -- & -6.66$^{+0.04}_{-0.05}${\footnotesize{($\pm 0.07$)}}{\footnotesize{($\pm 0.20$)}} & -- & -- & -7.09$^{+0.09}_{-0.11}${\footnotesize{($\pm 0.08$)}}{\footnotesize{($\pm 0.20$)}} \\
    %\RVSDSS{} [km/s] & 96.99$^{+10.37}_{-10.37}${\footnotesize{($\pm 10.00$)}}{\footnotesize{($\pm 14.41$)}} & 90.97$^{+11.61}_{-11.97}${\footnotesize{($\pm 10.00$)}}{\footnotesize{($\pm 15.46$)}} & 99.93$^{+15.36}_{-15.11}${\footnotesize{($\pm 10.00$)}}{\footnotesize{($\pm 18.22$)}} & 63.69$^{+6.08}_{-5.87}${\footnotesize{($\pm 10.00$)}}{\footnotesize{($\pm 11.65$)}} & 97.77$^{+11.66}_{-11.67}${\footnotesize{($\pm 10.00$)}}{\footnotesize{($\pm 15.36$)}} \\
    %\RVKeck{} [km/s] & 10.69$^{+6.40}_{-6.17}$ & 23.12$^{+1.28}_{-1.26}$ & 11.16$^{+1.78}_{-1.81}$ & 354.57$^{+3.31}_{-3.22}$ & 49.14$^{+2.10}_{-2.12}$ \\
    \AjitSDSS{} [--]  & 1.00$\pm$0.01 & 0.97$\pm$0.01 & 1.08$\pm$0.01 & 0.92$\pm$0.01 & 0.98$\pm$0.01 \\
    \AjitKeck{} [--]  & 1.25$\pm$0.01 & 1.44$\pm$0.01 & 1.09$\pm$0.01 & 1.08$\pm$0.01 & 1.11$\pm$0.01 \\
    \hline \hline
    Detections [no.] & 2 total (2 metals) & 6 total (5 metals) & 5 total (4 metals) & 4 total (3 metals) & 6 total (5 metals) \\
    \hline 
    \end{tabular}
    \vspace{0.1cm}
    %\begin{quote}
    %     \hspace{-0.17cm}\footnotesize{[\it{a}}]: \footnotesize{We note that \cecilia's results for \RVSDSS{} have not been corrected for the barycentric motion and gravitational redshift of the WDs. For this parameter, we assume a systematic error of 10~km/s based on the estimates provided by e.g. \citet{Abazajian:2009}.  } 
    %\end{quote}
\end{table*}

\section{Results} \label{sec:paper3_results}

\subsection{Bayesian Spectral Modelling with \cecilia} \label{sec:paper3_spectral_modelling}

The results of \cecilia's MCMCs are summarised in Table \ref{tab:paper3_mcmc_results}, together with their corresponding total uncertainties (\sigmatot). For each free model parameter, we assumed Gaussian and uncorrelated noise to approximate \sigmatot{} as the quadrature sum of its statistical (\sigmastat) and systematic error (\sigmasys). The former are directly obtained from the 1$\sigma$ confidence intervals of our MCMC and encapsulate the errors of the observations. In contrast, the latter arise from inaccuracies in \cecilia's ML predictions as well as from inherent imperfections in the WD atmosphere models used to train its networks. To ensure that \cecilia's uncertainties would not be greatly underestimated for our abundance parameters, we followed the methodology of BA24 and replaced their total error \sigmatot{} by a conservative noise floor of \sigmafloor{} whenever \sigmatot$<\sigma_{\mathrm{floor}}$. 

\begin{figure*}
      \centering
      \includegraphics[width=0.94\linewidth]{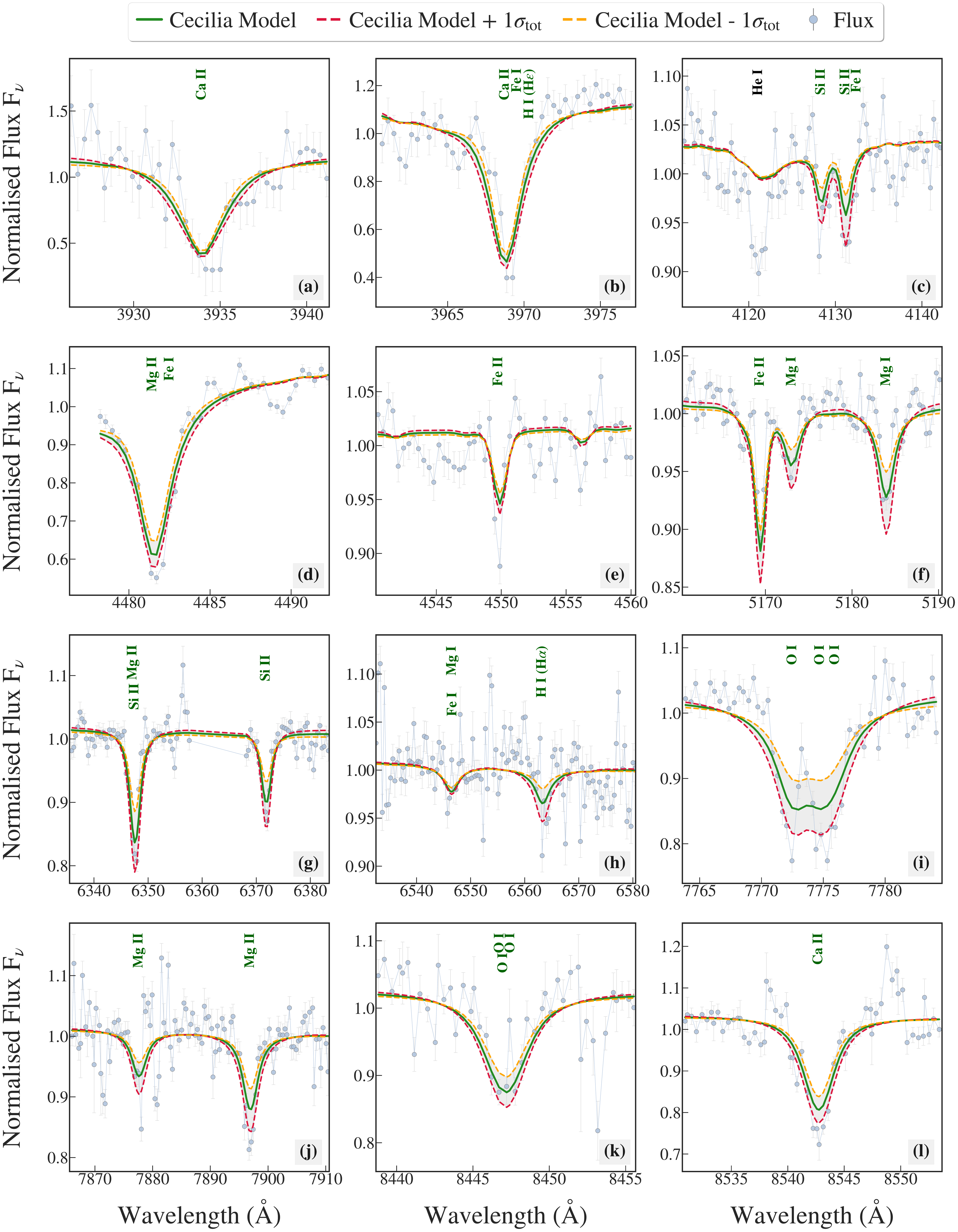}
      \caption{A selection of spectral windows showing \cecilia's best-fit RV-shifted MCMC model (in green) for the median-normalised \KeckESI{} spectrum of \wdzeroeightfivenine{} (in light blue; in air wavelengths). For reference, we also include \cecilia's predictions when modifying the abundances of the \textit{detected} elements by $\pm1$\sigmatot{} (red and orange). The green labels show all the detected elements, defined as those with \sigmastat{}$\leq$\sigmathresh{} and  at least one visible absorption line in the spectrum. We note that \cecilia{} struggles to model the depth of the Mg line at about 4,481~\angs{} (see panel \textit{d}). This is not the case for other Mg lines, which are fitted reasonably well by our code (e.g. panels \textit{g} and \textit{j}). Such behaviour may be an example of underestimated errors due to the high level of red (i.e. correlated) noise in panel \textit{d}. In Section \ref{sec:paper3_future_work}, we discuss how \cecilia{} can be improved to address this problem.}
      \label{fig:paper3_fit_wd0859}
\end{figure*}

\begin{figure*}
    \centering
    \includegraphics[width=0.78\linewidth]{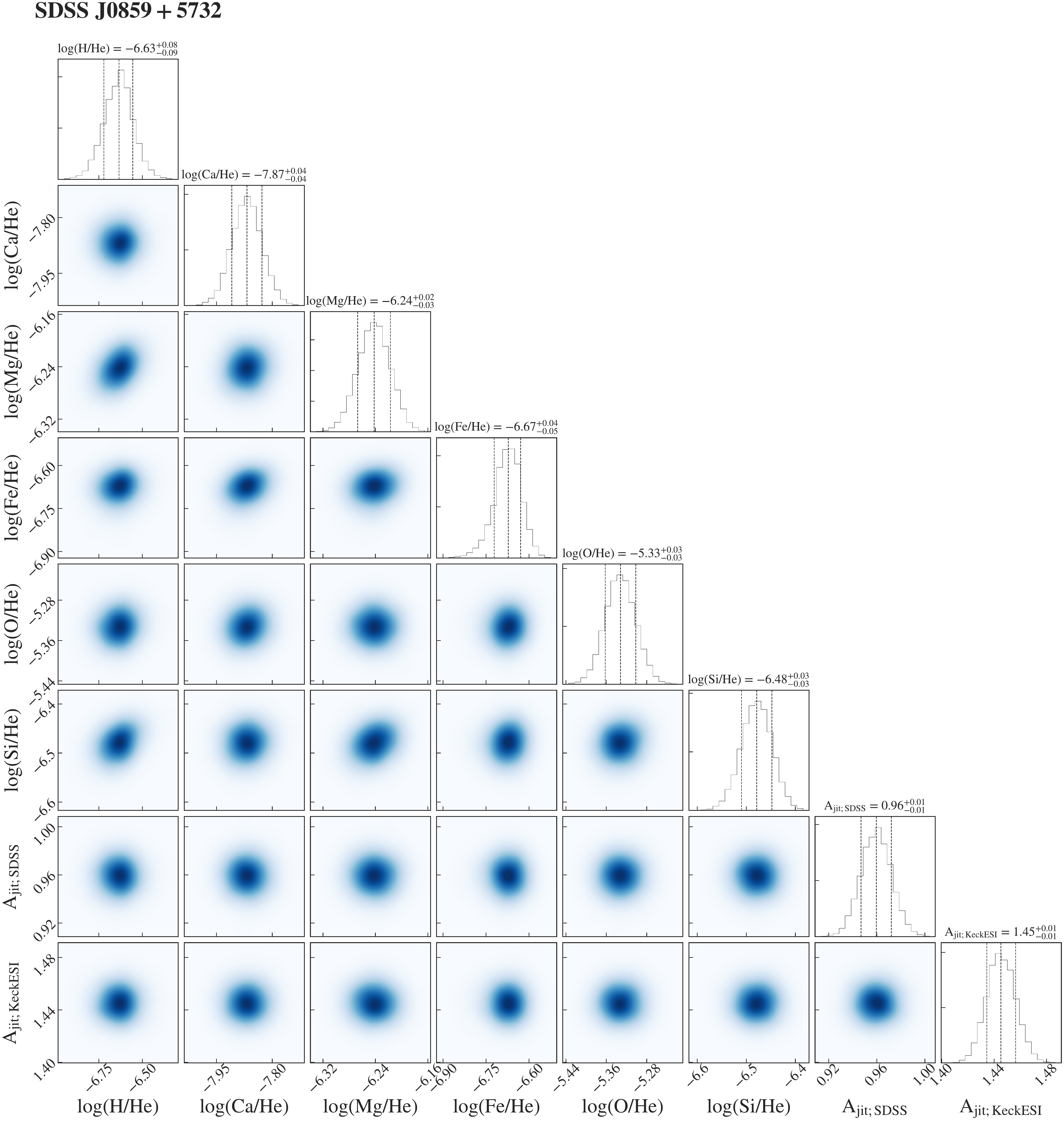}
       \caption{MCMC corner plot for \wdzeroeightfivenine. The off-diagonal plots illustrate the two-dimensional marginalised posterior distributions of the free model parameters, while the histogram panels along the diagonal show their one-dimensional marginalised distributions together with their median value and $1\sigma$ confidence interval. This Figure does not show our results for \cecilia's undetected elements. It also excludes the RV shifts because \cecilia{} does not correct for the barycentric motion and gravitational redshift of the white dwarf.}
       \label{fig:paper3_corner_wd0859}
\end{figure*}

The full spectroscopic solutions of \cecilia's optimisation routine are illustrated in \autoref{fig:paper3_wd_spectra}. In \autoref{fig:paper3_fit_wd0859} and Figures \ref{fig:paper3_fit_wd0231}-\ref{fig:paper3_fit_wd2311}, we provide zoomed-in panels of the \KeckESI{} best-fit models across different wavelength ranges. To visualise the quality of the model uncertainties, we also show \cecilia's predictions when modifying the abundances of the \textit{detected} elements by a factor of $\pm$1\sigmatot{}. We note that \cecilia's detections are primarily enabled by the quality of the \KeckESI{} observations. The \SDSS{} spectra are especially helpful for the analysis of the Ca II H$\&$K region between 3,930~\angs{} and 3,970~\angs{}, where the \KeckESI{} spectrograph has very limited sensitivity. However, they are too noisy elsewhere to contribute significantly to \cecilia's detections. Finally, \autoref{fig:paper3_corner_wd0859} and Figures \ref{fig:paper3_corner_wd0231}-\ref{fig:paper3_corner_wd2311} in the Appendix provide the corner plots of \cecilia's fits, demonstrating the convergence of our MCMCs and the lack of strong correlations between \cecilia's free model parameters.

 From \cecilia{}'s results, we can confidently identify a total of 2, 6, 5, 4, and 6 elements in the atmospheres of our targets (Table \ref{tab:paper3_mcmc_results}). In particular, we detect traces of Ca and Mg in \wdzerotwothreeone{}; H, O, Mg, Si, Ca, and Fe in \wdzeroeightfivenine{}; H, O, Mg, Si, and Ca in \wdoneonezeronine{}; H, O, Mg, and Ca in \wdonethreethreethree{}; and H, O, Mg, Si, Ca, and Fe in \wdtwothreeoneone. As discussed in Section \ref{sec:paper3_estimation_wd_abundances}, we consider these chemical species to be positive detections because their MCMC errors fall below our detectability cut-off (i.e. \sigmastat$<$\sigmathresh) and they all have at least one clear absorption line in the spectra. The most polluted WDs in our sample ---defined as those with the largest number of detected elements--- are \wdzeroeightfivenine{} and \wdtwothreeoneone{}, likely due to a combination of high accretion rates (see Table \ref{tab:paper3_table_props_detected_metals} and below) and relatively low effective temperatures. At lower \teff, the photospheres of WDs tend to be less opaque, which would make it easier for \cecilia{} to detect metal pollution.

Given that this paper represents the first application of \cecilia{} to the study of polluted WDs with no well-measured abundances, we chose to validate our results by fitting the spectra of our targets with the classical method of \citealt{Dufour:2012} (DF12). This comparison yielded consistent abundances within 0.2~dex, which is roughly the scatter we would expect due to methodological differences in the two fitting approaches (see Section \ref{sec:paper3_cecilia_overview}). More specifically, DF12 adopt a line-by-line $\chi^{2}$ minimisation approach, which effectively gives the same weight to each visible absorption feature. This differs from \cecilia's approach, which fits the full spectra simultaneously, assigning different weights to the lines based on their observational characteristics (e.g. line widths and strength, S/N).

\begin{figure*}
      \centering
      \includegraphics[width=1\linewidth]{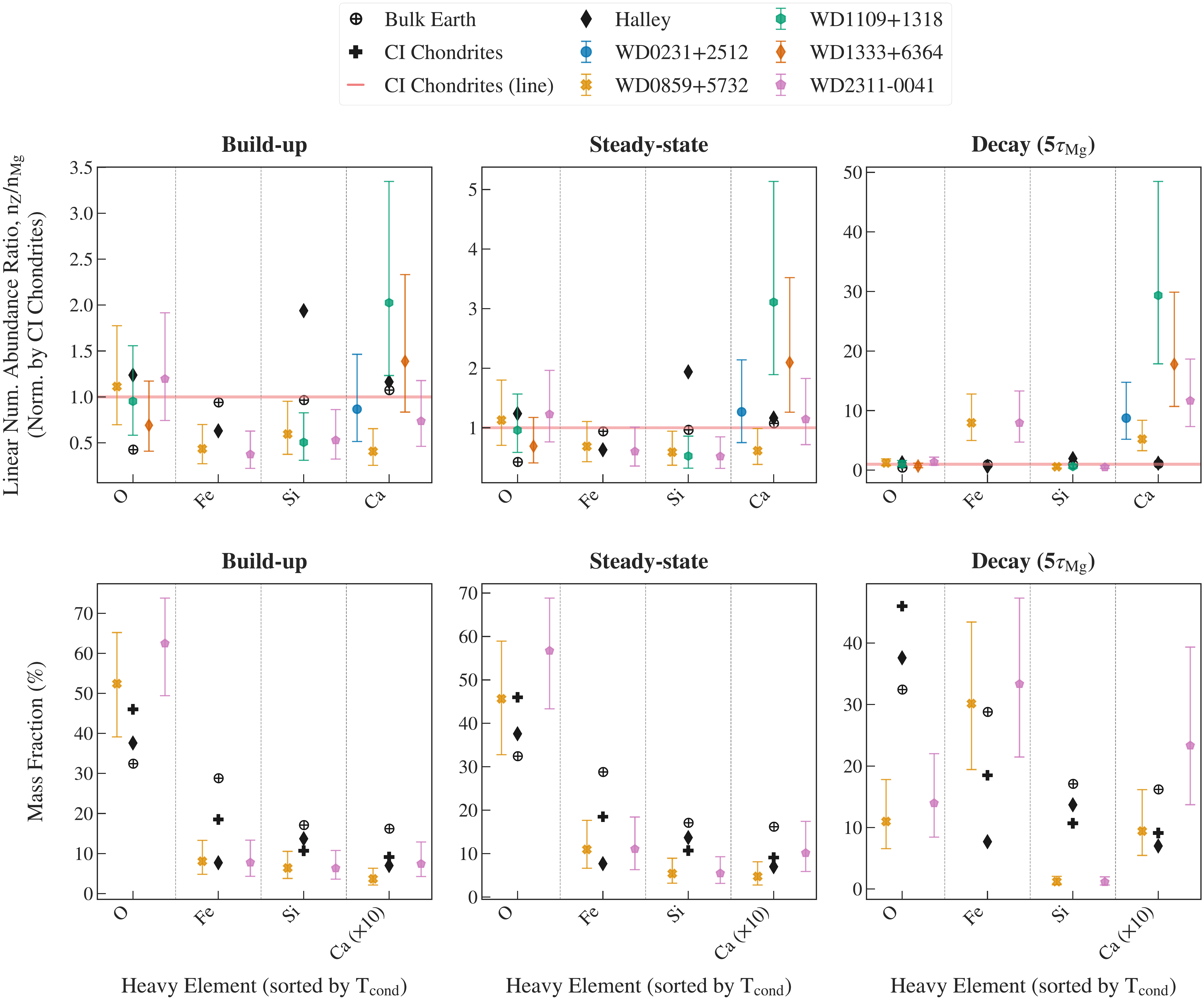}
      \caption{Compositional properties of the WD pollutants during build-up, steady-state, and decaying phase (first, second, and third columns, respectively). The top panels show the Mg-normalised linear abundance ratios of the accreted material relative to those of CI chondrites (red line; \citealt{Alexander:2019a_noncarbonaceous, Alexander:2019b_carbonaceous}). The bottom panels present the percent metal mass fractions for systems with simultaneous detections of at least the four major rock-forming elements (O, Fe, Si, Mg), with Ca values scaled by a factor of 10 for clarity. In each plot, we only include \cecilia's detected elements (excluding Mg, our reference metal), with error bars reflecting a total assumed abundance error of \sigmatot=\sigmafloor{}  (see Table \ref{tab:paper3_mcmc_results}). For comparison, we also show the properties of bulk Earth (black `$\oplus$' marker; \citealt{Allegre:2001}), comet Halley (black `$\Diamond$' marker; \citealt{Jessberger:1988}), and CI chondrites (black `$\bf{+}$' marker, bottom panels only). Heavy elements are sorted by increasing condensation temperature $T_{\rm cond}$ \citep{Lodders:2003}.}
      \label{fig:paper3_pollutant_props}
\end{figure*}

\subsection{Properties of Detected Metals} \label{sec:paper3_text_props_detected_metals}

For all five WDs, we implemented the methodology described in Section \ref{sec:paper3_estimation_wd_abundances} to estimate the mass ratio ($R_{\rm ratio, Z}$; Eq. \ref{eq:paper3_metal_mass_ratio}), atmospheric mass ($M_{\rm{Z}}$; Eq. \ref{eq:paper3_metal_mass}), and accretion rate ($\dot{M}_{\rm Z}$; Eq. \ref{eq:paper3_metal_accretion_rate}) of \cecilia's detected metals. Our results are summarised in Table \ref{tab:paper3_table_props_detected_metals}, together with lower limits on the total metal mass ($\sum M_{\rm{Z}}$) and total accretion rate ($\sum \dot{M}_{\rm Z}$) of each system.  For reference, we also include the logarithmic metal diffusion timescales ($\log_{10}\tau_{\rm{Z}}$) and the masses of the stellar convective layers (\Mcvz) obtained from the MWDD. 

Next, we used the equations presented in Section \ref{sec:paper3_estimation_pb_abundances} to infer the metal abundances of the WD pollutants. To this end, we focused on the build-up and steady-state phases of accretion, which yielded results more consistent with the enguflment of rocky exoplanetary debris than the decaying phase. For our calculations, we normalised \cecilia's best-fit abundances by Mg, that is, we set \nB$=n_{\rm Mg}$ and \tauB$= \tau_{\rm Mg}$ in Eq. \ref{eq:paper3_buildup}-\ref{eq:paper3_steady}. In Table \ref{tab:paper3_table_props_detected_metals}, we provide our normalised abundance measurements ($n_{\rm{Z}}/n_{\rm{Mg}}$) for \cecilia's detected elements. From these results, we also derived their corresponding percent mass fractions (MF$_{\rm Z}$; Eq. \ref{eq:paper3_metal_mass_fraction}). \autoref{fig:paper3_pollutant_props} illustrates our estimated abundances and MFs, together with those of bulk Earth \citep{Allegre:2001}, the CI chondrites \citep{Alexander:2019a_noncarbonaceous, Alexander:2019b_carbonaceous}, and comet Halley \citep{Jessberger:1988}. 

Considering the full range of abundance ratios, the compositions of the polluting bodies accreting onto the five WDs are largely consistent with those of CI chondrites to within 1-2\sigmatot. For \wdzeroeightfivenine{} and \wdoneonezeronine{}, their Ca/Mg ratios appear slightly lower and higher, respectively, than those of these primitive meteorites, although they are still consistent within 2\sigmatot. We also find that the O/Mg abundances of the four white dwarfs with oxygen detections (\wdzeroeightfivenine, \wdoneonezeronine, \wdonethreethreethree, \wdtwothreeoneone) are enhanced in comparison to that of rocky, bulk Earth-like material. From these results, we calculated the oxygen excess by assessing how much oxygen binds with the major elements in the accreted material.

% For \cecilia's undetected elements, we scaled their abundances relative to those of CI chondrites \citep{Lodders:2003} and used the resulting values for our calculations.
% \wdoneonezeronine{} is missing Al and Fe, and has an oxygen excess of 0.39$^{+0.22}_{-0.43}$ (0.91$\sigma$ with 82.3\% of samples giving a positive oxygen excess). \wdonethreethreethree{} has the lowest oxygen abundance ratio in the sample, and has an oxygen excess of just 0.08$^{+0.34}_{-0.65}$ (0.12$\sigma$ with 56.7\% of samples giving a positive oxygen excess), although given it is missing half of the rock forming elements (Al, Fe, Si), this value is only an estimate.}

The concept of oxygen excess, first introduced by \citet{Klein:2010}, refers to the number of oxygen atoms that remain unaccounted for after calculating the amount of oxygen needed to form common rock-forming oxides (CaO, SiO$_2$, FeO, MgO). In this work, we investigated this phenomenon for \wdzeroeightfivenine{} and \wdtwothreeoneone{}, which exhibit the four main rock-forming elements in their spectra (O, Fe, Si, Mg). Assuming steady-state accretion and that iron is in the form of FeO, we followed the methodology of \citet{Rogers:2024b} to quantify the oxygen excess and its significance. In particular, we employed Monte Carlo techniques to sample plausible oyxgen excess values given our assumed total error on \cecilia's predicted abundances. Our analysis indicates that \wdzeroeightfivenine{} and \wdtwothreeoneone{} have oxygen excesses of 0.64$^{+0.13}_{-0.25}$ and 0.68$^{+0.12}_{-0.23}$, respectively. Accounting for the asymmetry of the Monte Carlo sampled values, we  calculated the fraction of sampled values showing a positive oxygen excess, obtaining 95.5$\%$ and 96.4$\%$ for \wdzeroeightfivenine{} and \wdtwothreeoneone{}, respectively. Comparing these probabilities to the 2$\sigma$ significance threshold of \citealt{Brouwers:2022} (equivalent to 95.45$\%$ confidence),\footnote{As discussed in \citet{Brouwers:2022}, the use of 2$\sigma$ confidence threshold is a conservative trade-off between minimising false positives (which would become abundant at the 1$\sigma$ level), and the limitations imposed by errors in the stellar abundances and accretion rates (which would make the 3$\sigma$ very restrictive). Our methodology follows the same conservative approach, but assumes a steady-state accretion model.} we find that both targets have statistically significant oxygen excesses. Our results are also similar to those reported by \citet{Brouwers:2022} for GALEXJ2339 (0.61, using data from \citealt{Klein:2021}), and WD1232$+$563 (0.57, using data from \citealt{Xu:2019}), where H$_2$O is considered the most likely carrier of excess oxygen. It should be noted that the derived H abundances for \wdzeroeightfivenine{} and \wdtwothreeoneone{} are too low to account for the oxygen excess, so it is likely additional oxygen carriers, such as carbon-bearing species, would be required to explain the observed stellar abundances.

Based on \cecilia's metal detections, our work suggests that the total metallic mass content of the WD pollutants ranges from $5.43 \times 10^{21}$~g (for \wdzerotwothreeone) to $1.92 \times 10^{23}$~g (for \wdzeroeightfivenine) ---in agreement with previous studies of metal pollution \citep[e.g.][]{Zuckerman:2010, Farihi:2016, Xu:2019}. At the lower end, our inferred mass is comparable to that of a small- or medium-sized asteroid in the Solar System. For example, 52~Europa has a mass of approximately $2.4 \times 10^{22}$~g, slightly above our lower bound. At the upper end, our result is comparable to the accretion of a much larger body, similar to Saturn's icy moon Enceladus ($1.08 \times 10^{23}$~g; \citealt{Jacobson:2022}), or the large asteroids Vesta ($2.5 \times 10^{23}$~g; \citealt{Russell:2012}) and Pallas ($2.05 \times 10^{23}$~g; \citealt{Carry:2012}). Assuming a typical density of 2.2~$\rm g/cm^{3}$ for CI chondrites \citep{Britt:2004}, the total accreted masses derived in this work would result in a single pollutant with diameters between approximately 170-550~km. Lastly, in terms of total accretion rate, our results range between $10^{7} - 10^{9}$~g/s, which is consistent with the observed distribution for He-atmosphere polluted WDs with similar effective temperatures \citep{Rogers:2024a}.

\section{Discussion} \label{sec:paper3_discussion} 

\subsection{Limitations and Insights of Compositional Analysis}\label{sec:paper3_discussion_limitations} 

Our geological inferences are inherently limited by the number of detected metals in the WDs and our conservative noise floor of 0.20~dex on the stellar abundances. Another limitation is that these systems are too faint to be observed with current UV facilities ---which would typically allow for the detection of more elements, including volatiles such as C, N, or P---, so our conclusions are restricted to the optical spectra considered in this work. Recognising these observational caveats, the five WDs in our sample appear to be accreting rocky extrasolar material with a composition similar to those of CI chondrites. These results should be interpreted with caution, as the absence of major detected elements like Fe, Si, and Al introduces uncertainties in our calculations, especially when assuming chondritic values for some of the primary rock-forming elements. In particular, Fe and Si are critical for the formation of planetary mantles and cores, so their absence in some WDs limits our ability to make reliable assumptions about the geochemistry and internal structure of the accreted debris.

With respect to \wdzeroeightfivenine{} and \wdtwothreeoneone{}, both systems show enhanced oyxgen levels compared to bulk Earth, with significant oxygen excesses ($>2\sigma$) pointing to the accretion of oxygen-rich exoplanetary material. An important caveat is that the assumed oxidation state of Fe can influence the magnitude of the inferred oxygen excess, as Fe could be in the form of metallic Fe, FeO, or Fe$_2$O$_3$. If the Fe were instead fully or partially in metallic form, the inferred oxygen excess would increase; conversely, if all Fe were assumed to be present as Fe$_2$O$_3$, the resulting oxygen excesses would be 0.60$^{+0.14}_{-0.28}$ for \wdzeroeightfivenine{} and 0.65$^{+0.13}_{-0.25}$ and \wdtwothreeoneone{}. In these cases, 94.1$\%$ and 95.6$\%$ of the Monte Carlo samples lie above zero, indicating a positive oxygen excess. When compared to the 2$\sigma$ significance threshold of \citealt{Brouwers:2022} (corresponding to 95.45$\%$ confidence), the oxygen excess in \wdtwothreeoneone{} would remain statistically significant, while the result for \wdzeroeightfivenine{} would fall just below the threshold, offering tentative evidence for an oxygen excess even under this conservative oxidation scenario. Another potential source of uncertainty in our oxygen excess calculations is the phase of accretion, with build-up making the excess more pronounced, and declining making it less significant. The detection of infrared flux excesses would offer independent evidence of ongoing accretion \citep{Bonsor:2017}, but none of our targets have existing \textit{WISE} and \textit{Spitzer} photometry to support this hypothesis. Despite this limitation, the predicted abundances of the polluting bodies resemble those of rocky material from the Solar System assuming build-up or steady-state accretion, which would make it unlikely that these WDs are accreting in the declining phase. Furthermore, \wdzeroeightfivenine{} and \wdtwothreeoneone{} retain their oxygen excess up to 4.8 and 5 times the Mg sinking timescale into the declining phase, respectively. At these times, the abundance patterns of their accreted material deviate significantly from those of known rocky Solar System bodies (see \autoref{fig:paper3_pollutant_props} and \autoref{fig:paper3_decay_evolution}), which would further support our interpretation that these systems are not in the declining phase and that their oxygen excesses are real. Therefore, our results add two more WDs to the growing smaple of systems with notable oxygen excesses \citep{Farihi:2011, Farihi:2013, Raddi:2015, Xu:2017, Hoskin:2020, Klein:2021, Rogers:2024a}, demonstrating that oxygen-rich material can survive during post-MS evolution and subsequently accrete onto WDs.

An additional insight from our geological analysis is the tentative evidence for Si depletion relative to bulk Earth and CI chondrites (see \autoref{fig:paper3_pollutant_props}). For example, our estimated steady-state Mg/Si ratios are approximately 1.5, 1.7, and 1.7 for \wdzeroeightfivenine{}, \wdoneonezeronine, and \wdtwothreeoneone{}, respectively (see Table \ref{tab:paper3_table_props_detected_metals}). Although these values have relatively large uncertainties, they appear to exceed Earth's mantle ratio of about 1.3 \citep{Ringwood:1989}. Therefore, if real, they would point to the engulfment of polluting bodies with a mineralogy primarily composed of magnesium silicates ---dominated by olivine with lesser amounts of pyroxene---, and relatively depleted in silicon compared to bulk Earth. Together, the high Mg/Si ratios and oxygen excesses observed in \wdzeroeightfivenine{} and \wdtwothreeoneone{} would imply the accretion of oxygen-rich exoplanetary material with significant amounts of magnesium silicates, such as olivine. 

Beyond individual systems, an elevated Mg/Si ratio could have important consequences for extrasolar geochemistry, shaping factors such as water storage capacity, magnetic fields, or tectonic activity. For instance, high Mg/Si ratios would favour the formation of forsterite-rich olivine, which is less dense and has a lower viscosity than pyroxene \citep{Mackwell:1991, Hansen:2015}. This difference in viscosity would impact the thermal and dynamical evolution of the mantle; in particular, lower-viscosity olivine would lead to more vigorous mantle convection and faster cooling of the core, which could potentially shorten the lifetimes of magnetic fields and plate tectonics  \citep{Spaargaren:2020, Spaargaren:2023} ---two critical processes for sustaining long-term planetary habitability \citep{Stern2024}.

\subsection{\cecilia's Current Performance and Future Work} \label{sec:paper3_future_work}

In this paper, we have presented several  upgrades to \cecilia{}, including the implementation of a more complex MCMC log-likelihood function to allow for joint spectroscopic fits with a more careful treatment of noise and spectral lines  (see Section \ref{sec:paper3_cecilia_upgrades}). We have also reconfigured \cecilia's MCMC to enable the calculation of upper abundance limits for those elements with unclear or unobservable absorption lines. Although these upgrades have had a positive impact on \cecilia's performance, there are still many opportunities to improve our code. For instance, as discussed in BA24, \cecilia{} could be retrained with more WD models featuring synthetic photometric observations, lower and higher abundance ranges, additional heavy elements (e.g. Al, Na), and  new regions of the spectrum, such as the UV. 

Building upon the recommendations of BA24, we propose several new directions for future work to further enhance \cecilia's  capabilities. Our suggestions can be divided into two broad categories: data processing, and Bayesian inference techniques. In relation to the former, we would like to improve our methodology for correcting telluric contamination by Earth's atmosphere, particularly redward of about 6,000~\angs. In this paper, we cautiously removed all the most important telluric bands from our \KeckESI{} spectra, even if some of these bands contained potential metal absorption lines (see Section \ref{sec:paper3_data_treatment}). To address this limitation, we aim to develop a more selective and targeted approach to removing telluric lines in ground-based optical observations. For example, we could correct an observed WD spectrum using a telluric absorption model generated by packages such as the Fortran  line-by-line radiative transfer code \texttt{LBLRTM}  \citep{Clough:2005}\footnote{\url{http://rtweb.aer.com/lblrtm.html}} or its Python \texttt{telfit} implementation\footnote{\url{https://telfit.readthedocs.io/en/latest/}} \citep{Gullikson:2013_telfit, Gullikson:2014_telfit}. The latter uses  a non-linear least-squares Levenverg-Marquardt algorithm to model  telluric effects, given a list of molecular line strengths from the HITRAN database and a nighttime atmosphere model containing the pressure, temperature, and abundances of different molecules as a function of height. We note that \texttt{telfit} can also employ custom atmosphere profiles at the specific location of the user's observatory. 

%bayesian things
In addition to developing better data processing techniques, we have also identified multiple opportunities to improve \cecilia's optimisation procedure, which we summarise below. 

\begin{enumerate}[noitemsep,topsep=1pt,leftmargin=14pt,listparindent=1.5em]
    \vspace{5pt}
    \item \textit{Use of Gaussian Processes}: Given the inaccuracy of state-of-the-art atmosphere models to account for the problem of neutral He line broadening from atom collisions, we would like to explore the use of Gaussian Process (GP) regression to improve the quality of \cecilia's results, especially around poorly-modelled absorption lines. Broadly speaking, GPs are powerful mathematical tools to perform joint non-parametric\footnote{Parametric models have a fixed number of unknown parameters. This differs from non-parametric models, which can have an arbitrarily large number of dimensions.} fits to astronomical observations with both uncorrelated (i.e. \textit{white}) and correlated (i.e. \textit{red}) sources of noise (e.g. instrumental effects or stellar variability). Mathematically, GPs are defined as generalisations of multivariate Gaussian distributions with a covariance matrix (also known as \textit{kernel} function) that encapsulates the underlying stochastic (i.e. random) correlation structures between adjacent datapoints \citep{Rasmussen:2006}. In the context of \cecilia, we could implement a GP noise model to account for imperfections in the atmosphere models of He-rich polluted WDs. This approach would downweight problematic spectral regions by inflating the systematic errors of poorly fitted absorption lines, hence penalising atmospheric solutions with under- or overestimated abundances. On a practical level, the use of GPs would involve balancing \cecilia's model flexibility, uncertainty estimates, and computational speed (\citealt{Czekala:2015}, C15). On the one hand, GPs may be more efficient at capturing underestimated systematics in the WD atmosphere models, potentially leading to more realistic and conservative results. On the other, they may inflate the errors on the retrieved model parameters, while also slowing down \cecilia's optimisation routine. These trade-offs may nonetheless be reasonable if they ultimately improve the reliability and accuracy of our code. In the future, we can explore the integration of GPs with publicly available software packages, such as \texttt{tinyGP} \citep{ForemanMackey:2024_tinyGP} or \texttt{starfish} (C15).\footnote{These packages can be accessed via \url{https://github.com/dfm/tinygp} and \url{https://starfish.readthedocs.io/en/latest/index.html}, respectively.}

    \vspace{5pt}
    \item \textit{Exploration of new sampling mechanisms}: The main optimisation method employed by \cecilia{} is the differential evolution MCMC algorithm of \citet{Ter:2006}. As we seek to improve the efficiency of our pipeline, we could also experiment with different inference techniques, such as nested sampling \citep{Skilling:2004, Skilling:2006}. Unlike MCMCs, which are designed to sample the posterior distribution  directly from the likelihood function and the prior density, nested sampling revolves around estimating the Bayesian evidence of a model. This parameter is very difficult to determine, so nested sampling addresses this problem in a dynamical and iterative way. More specifically, the algorithm draws an ensemble of random ``live points'' from the prior, removes the point with the lowest likelihood, generates new live points with higher likelihoods, and repeats this process  until the Bayesian evidence of a model satisfies a certain threshold. Therefore, unlike MCMCs, which are ``memoryless'' systems in which the behaviour of two walkers only depends on their previous state, nested sampling methods systematically explore a large volume of the parameter space and gradually compress it based on regions of higher probability. With this strategy, they are more robust to poor initial guesses as well as to multi-modal distributions between different parameters \citep{Ashton:2022}. A potential nested sampling implementation is the open-source package \texttt{dynesty} \citep{Speagle:2020, Koposov:2023}.\footnote{\url{https://dynesty.readthedocs.io/en/latest/dynamic.html}}
     %or simulated annealing
   
    \vspace{5pt}
    \item \textit{Bayesian model comparison}: Central to our understanding of metal pollution is our degree of belief in spectroscopic fits of polluted WD spectra, especially when obtained with ML codes that may not often be easily explainable. In the future, we could improve \cecilia{} by integrating a Bayesian model comparison framework to assess the relative probability of different ML predictions in the absence or presence of a heavy element. This can be achieved by calculating the so-called ``Bayes factor,'' which is defined as the ratio 
    \begin{equation}
    R \equiv \frac{p_{M_{1}} \cdot \pi_{M_{1}}}{p_{M_{2}} \cdot \pi_{M_{2}}},
    \label{eq:paper3_bayes}
    \end{equation}
    where $M_{1}$ and $M_{2}$ are two competing models, $p_{M_{1}}$ and $p_{M_{2}}$ are their posterior probability distributions given the observed data, and $\pi_{M_{1}}$ and $\pi_{M_{2}}$ are our \textit{a priori} beliefs on each model. From this expression, R$\gg$1 and B$\approx$0 would indicate a strong preference for $M_{1}$ and $M_{2}$, respectively. However, more complex interpretations are also possible, such as those of \citet{Jeffreys:1998} and \citet{Kass:1995}. In general, Eq. \ref{eq:paper3_bayes} is difficult to solve, so there are several approximate forms of Bayesian model comparison. For example, the so-called Bayesian Information Criterion (BIC) evaluates the accuracy and complexity of different models based on the maximum value of their likelihood function ($\hat{\mathcal{L}}$), the number of points in the observations ($n$), and their number of free model parameters ($k$), i.e., $\rm{BIC}=-2\ln(\hat{\mathcal{L}})+k\ln(n)$.
   
    %\item \textit{Alternative methods to bayesian optimisation}: A key driver of \cecilia's design was the need to obtain robust uncertainty estimates for our model parameters. For this purpose, we chose to implement an MCMC in the framework of Bayesian probabilistic inference. Nonetheless, we could also explore other optimisation techniques, such as Genetic Algorithms (GAs). These GAs are powerful ``evolutionary'' methods  inspired by the princples of genetics and natural selection \citep{Vie:2020}. Due to their high versatility, they are often capable of identifying global optimal solutions in complex, non-linear problems. Moreover, they do not require the computation of derivatives (i.e. they are gradient-free algorithms). A simple GA implementation may not necessarily generate model uncertainties (\blue{ref}), but we could explore the possiblity of implementing a GA-based optimisation procedure with an \textit{adhoc} error quantification procedure.  

    \vspace{5pt}
    \item \textit{Improved Treatment of Resolving Power or  Spectral Resolution}: Our analysis of \SDSS{} observations assumes a constant resolving power of R=2,000 between about 3,800\angs{} to 9,000\angs{}, even though $R$ changes  significantly by about 67$\%$ (from 1,500 to 2,500) across this wide wavelength range. Our simplification does not fully capture the intrinsic properties of the spectra, so future iterations of \cecilia{} could adopt a more adaptive approach, either by convolving the data to a constant linear resolution in wavelength space, or by directly supporting a variable resolving power or resolution to better preserve the native characteristics of the observations. 
  
    \vspace{5pt}
    \item \textit{Automated detection of chemical elements}: In this paper, we consider an element to be detected if (i) its \cecilia{} predicted abundance had a statistical error lower than or equal to a conservative detectability threshold of \sigmathresh, and (ii) if it exhibited at least a clear absorption line in the spectrum. This validation process could be automated in order to streamline all the criteria associated to a robust detection.  
\end{enumerate}

\section{Conclusions}\label{sec:paper3_conclusions} 

In this paper, we have used the ML-pipeline \cecilia{} to constrain the physical and chemical properties of five metal polluted He-atmosphere  WDs. We started by performed a joint fit to their \SDSS{} (R=2,000) and \KeckESI{} (R=4,500) spectra with an MCMC model consisting of 17 parameters (\teff, \logg, 11 elemental abundances, and a jitter term and RV shift per spectrum). We then estimated the geological composition of the WD pollutants from \cecilia's predicted stellar abundances. The main computational and scientific conclusions of our work are summarised below.

\vspace{5pt}
\begin{enumerate}[noitemsep,topsep=1pt,leftmargin=26pt]
    \item \textit{Upgrades to \cecilia{} (Section \ref{sec:paper3_cecilia_upgrades}}): We have improved \cecilia's  fitting procedure in two ways. First, we have re-parametrised its MCMC to sample the posterior distribution of the model parameters in linear space. This change will eventually allow us to calculate robust upper abundance limits for those elements with undetected spectral lines. Second, we have modified our log-likelihood function (Eq. \ref{eq:paper3_mcmc_loglike}) to allow for a joint fit of multiple spectra, regardless of their observational characteristics. We have also incorporated a jitter term per spectrum to model unaccounted sources of noise in the observations or in \cecilia's training atmosphere models. 
    \vspace{5pt}
    
    \item \textit{Atmospheric Analysis (Section \ref{sec:paper3_estimation_wd_abundances}}): We have identified a total of 2, 6, 5, 4, and 6 chemical elements in the atmospheres of \wdzerotwothreeone{} (Mg, Ca), \wdzeroeightfivenine{} (H, O, Mg, Si, Ca, Fe), \wdoneonezeronine{} (H, O, Mg, Si, Ca), \wdonethreethreethree{} (H, O, Mg, Ca), and \wdtwothreeoneone{} (H, O, Mg, Si, Ca, Fe), respectively. In terms of detected metals, \wdzeroeightfivenine{} and \wdtwothreeoneone{} are the most heavily polluted in our sample. For all five WDs, \cecilia{} constrains the abundances of the detected elements with greater precision than our assumed noise floor of $\leq$0.20 dex ---a performance comparable to that of classical WD fitting methods. We emphasise, however, that \cecilia{}'s predictive power is fundamentally restricted by the quality of its training models and by its ML interpolation errors (see BA24 for a detailed discussion of systematics using synthetic data). In other words, although \cecilia{} may yield lower statistical uncertainties than conventional techniques, its overall performance cannot exceed that of classical methods because it is inherently limited by the accuracy of its underlying WD atmosphere models.
   \vspace{5pt}

    \item \textit{Geological Analysis (Sections \ref{sec:paper3_estimation_pb_abundances}-\ref{sec:paper3_text_props_detected_metals}}): Our results indicate that the five WDs engulfed rocky extrasolar material with a bulk composition largely consistent with those of CI chondrites (within 1-2\sigmatot), with the main rock-forming elements (Mg, Fe, Si, and O) accounting for most of their interior composition by mass. Among these systems,  \wdzeroeightfivenine{} and \wdtwothreeoneone{} have statistically significant ($>2\sigma$) oxygen excesses, which could indicate the accretion of oxygen-rich extrasolar material. These findings are not conclusive given the limited number of detected metals in each WD and the uncertainties in \cecilia's stellar abundances. However, even within this margin of error, our analysis is sensitive enough to identify compositional deviations in the WD pollutants. In the coming years, as next-generation optical/UV telescopes become operational, it may be possible to disentangle their full geochemical properties and place them in the broader context of extrasolar  compositions.
\end{enumerate}
\vspace{5pt}

To conclude, we have demonstrated that \cecilia{} can constrain the elemental abundances of He-rich polluted WDs with no \textit{a priori} knowledge of their atmospheric composition and with minimal human supervision. As we venture into the era of massively multiplexed observational surveys (e.g. DESI, SDSS-V, WEAVE, 4MOST), \cecilia{} can be used to rapidly analyse large volumes of data within a practical amount of time, therefore offering a solution to the human-in-the-loop problem of conventional WD characterisation methods. In doing so, \cecilia{}  aims to lay the foundations for a methodological shift towards population-wide studies of metal pollution ---a ``Big Data'' revolution with the potential to offer new insights into the  properties of extrasolar worlds.

\section*{Acknowledgements}

%personal
MBA thanks Oriol Abril-Pla for his valuable feedback on this work. SB acknowledges support from the Banting Postdoctoral Fellowship and CITA National Fellowship programs. SX is supported by NOIRLab, which is managed by the Association of Universities for Research in Astronomy (AURA) under a cooperative agreement with the National Science Foundation. LKR acknowledges support of an ESA Co-Sponsored Research Agreement No. 4000138341/22/NL/GLC/my = Tracing the Geology of Exoplanets. MBA, AB, LKR acknowledge support of a Royal Society University Research Fellowship, URF\textbackslash R1\textbackslash 211421. At MIT, MBA was supported by the MIT Department of the Earth, Atmospheric, and Planetary Sciences, NASA grants 80NSSC22K1067 and 80NSSC22K0848, and the MIT William Asbjornsen Albert Memorial Fellowship.

%Gaia, SDSS, MWDD, SIMBAD, Vizier
This work has made use of data from the European Space Agency (ESA) mission \Gaia{} (\url{https://www.cosmos.esa.int/gaia}), processed by the \Gaia{} Data Processing and Analysis Consortium (DPAC,
\url{https://www.cosmos.esa.int/web/gaia/dpac/consortium}). Funding for the DPAC has been provided by national institutions, in particular the institutions participating in the \Gaia{} Multilateral Agreement. It has also used the Sloan Digital Sky Survey (SDSS), which is a joint project of The University of Chicago, Fermi National Accelerator Laboratory, The Institute for Advanced Study, The Japan Participation Group, The Johns Hopkins University, The Max-Planck-Institute for Astronomy (MPIA), The Max-Planck-Institute for Astrophysics (MPA), New Mexico State University, Princeton University, The United States Naval Observatory, The University of Washington, Los Alamos National Laboratory, and The University of Pittsburgh. Apache Point Observatory, site of the SDSS telescopes, is operated by the Astrophysical Research Consortium (ARC). Finally, this work has made use of public data from the Montreal White Dwarf Database, the SIMBAD database and VizieR catalogue (both operated at CDS, Strasbourg, France)

%Keck
Some of the data presented herein were obtained at Keck Observatory, which is a private 501(c)3 non-profit organization operated as a scientific partnership among the California Institute of Technology, the University of California, and the National Aeronautics and Space Administration. The Observatory was made possible by the generous financial support of the W. M. Keck Foundation. The authors wish to recognize and acknowledge the very significant cultural role and reverence that the summit of Maunakea has always had within the Native Hawaiian community. We are most fortunate to have the opportunity to conduct observations from this mountain.

%Python
This work has employed the following open-source software packages: \texttt{Python} \citep{Python}, \texttt{numpy} \citep{numpy}, \texttt{scipy} \citep{scipy}, \linebreak 
\texttt{matplotlib} \citep{matplotlib}, \texttt{astropy} \citep{astropy:2018}, \texttt{pandas} \citep{pandas:2010}, \citep{Markwardt:2009}, \texttt{edmcmc} \citep{Vanderburg:2021_edmcmc}, \texttt{tensorflow} \citep{tensorflow2015-whitepaper}, and \texttt{corner} \citep{corner, corner_luger}.

%%%%%%%%%%%%%%%%%%%%%%%%%%%%%%%%%%%%%%%%%%%%%%%%%%
\section*{Data Availability}

The \SDSS{} and \KeckESI{} spectra of our targets can be downloaded from the \SDSS{} DR18 and the \emph{Keck} online databases. The spectra of the standard stars described in Section \ref{sec:paper3_data} can be obtained from the STScI archive. 

%%%%%%%%%%%%%%%%%%%% REFERENCES %%%%%%%%%%%%%%%%%%

% The best way to enter references is to use BibTeX:

\bibliographystyle{mnras}
\bibliography{ref} 

%%%%%%%%%%%%%%%%%%%%%%%%%%%%%%%%%%%%%%%%%%%%%%%%%%

%%%%%%%%%%%%%%%%% APPENDICES %%%%%%%%%%%%%%%%%%%%%

\appendix \label{sec:paper3_appendix}

\section{Figures}

\begin{figure*}
    \centering
    \includegraphics[width=0.99\linewidth]{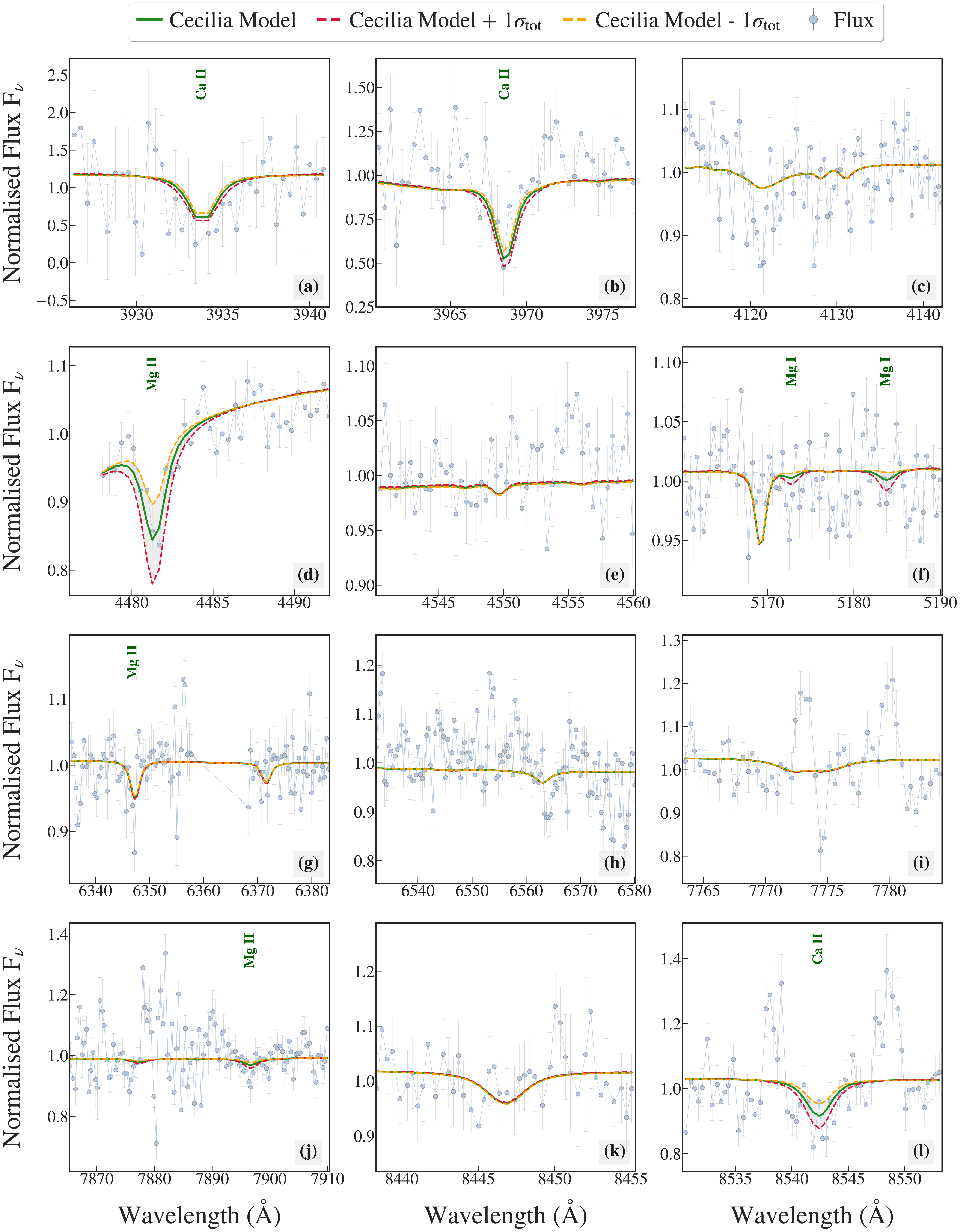}
    \caption{\cecilia's best-fit MCMC model (in green)  for the median-normalised \KeckESI{} spectrum of \wdzerotwothreeone, with its corresponding $\pm$1\sigmatot{} models (in red and orange).}
    \label{fig:paper3_fit_wd0231}
\end{figure*}

\begin{figure*}
      \centering
      \includegraphics[width=0.99\linewidth]{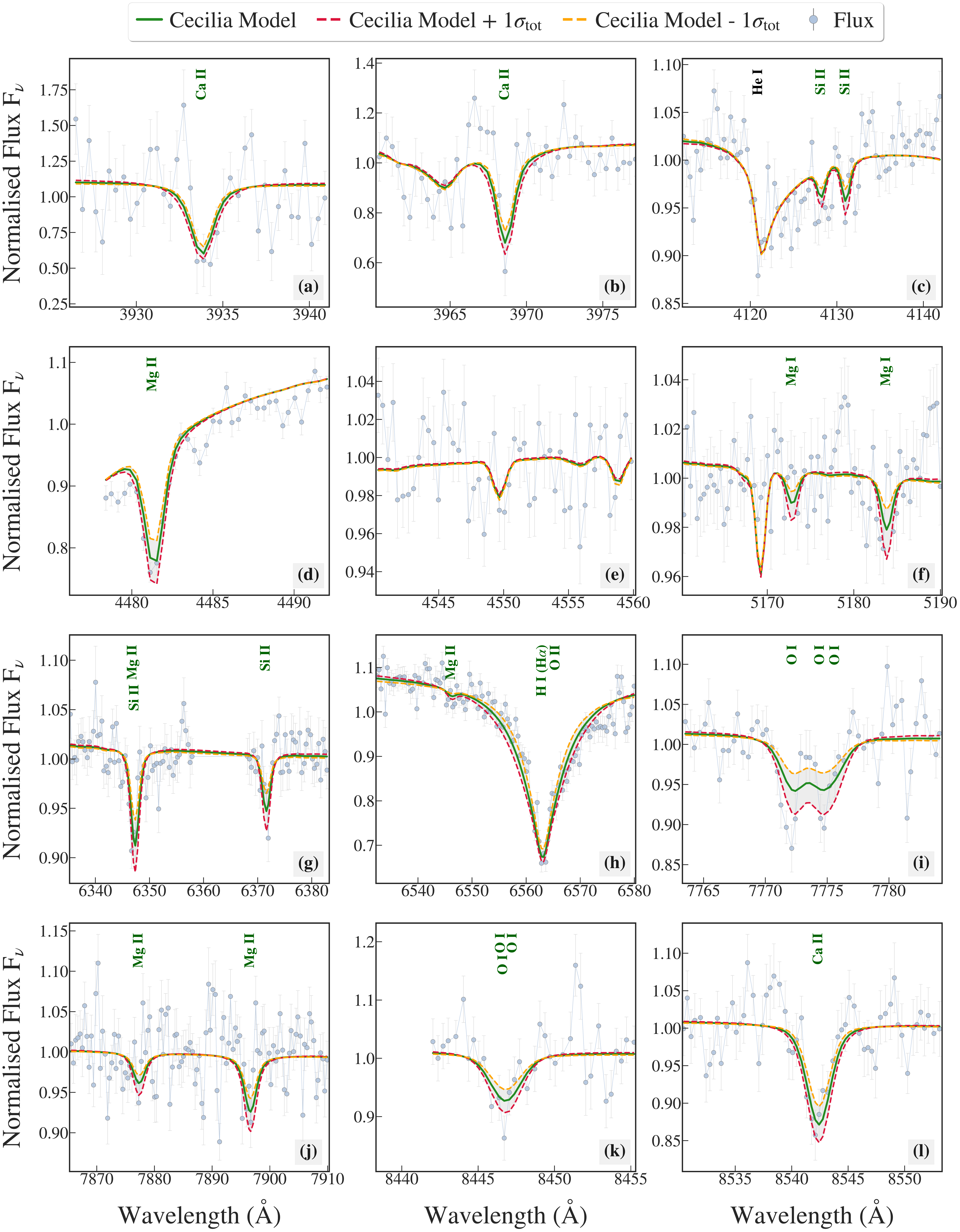}
      \caption{\cecilia's best-fit MCMC model (in green)  for the median-normalised \KeckESI{} spectrum of \wdoneonezeronine, with its corresponding $\pm$1\sigmatot{} models (in red and orange).}
      \label{fig:paper3_fit_wd1109}
\end{figure*}

\begin{figure*}
      \centering
      \includegraphics[width=0.99\linewidth]{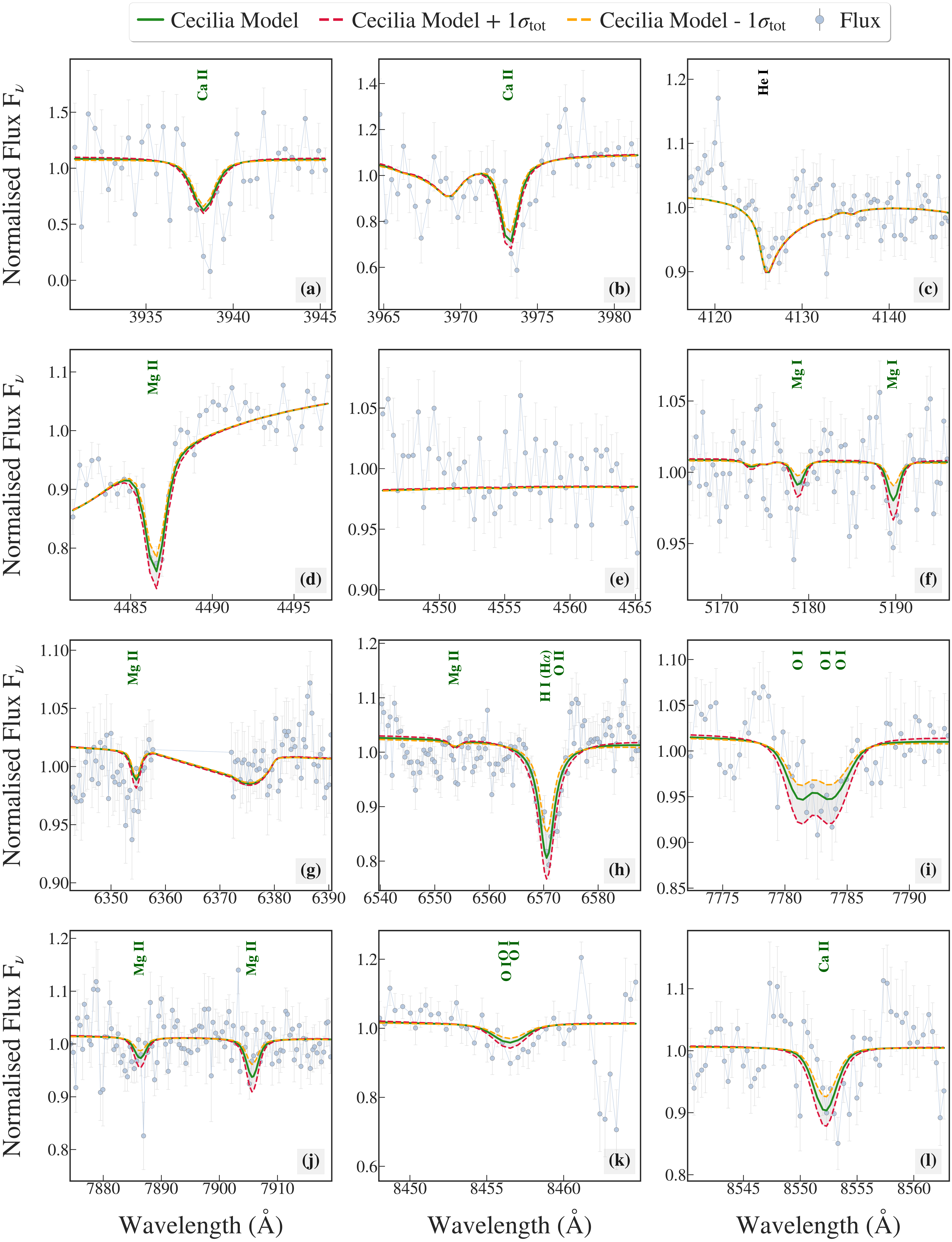}
      \caption{\cecilia's best-fit MCMC model (in green)  for the median-normalised \KeckESI{} spectrum of \wdonethreethreethree, with its corresponding $\pm$1\sigmatot{} models (in red and orange).}
      \label{fig:paper3_fit_wd1333}
\end{figure*}

\begin{figure*}
      \centering
      \includegraphics[width=0.99\linewidth]{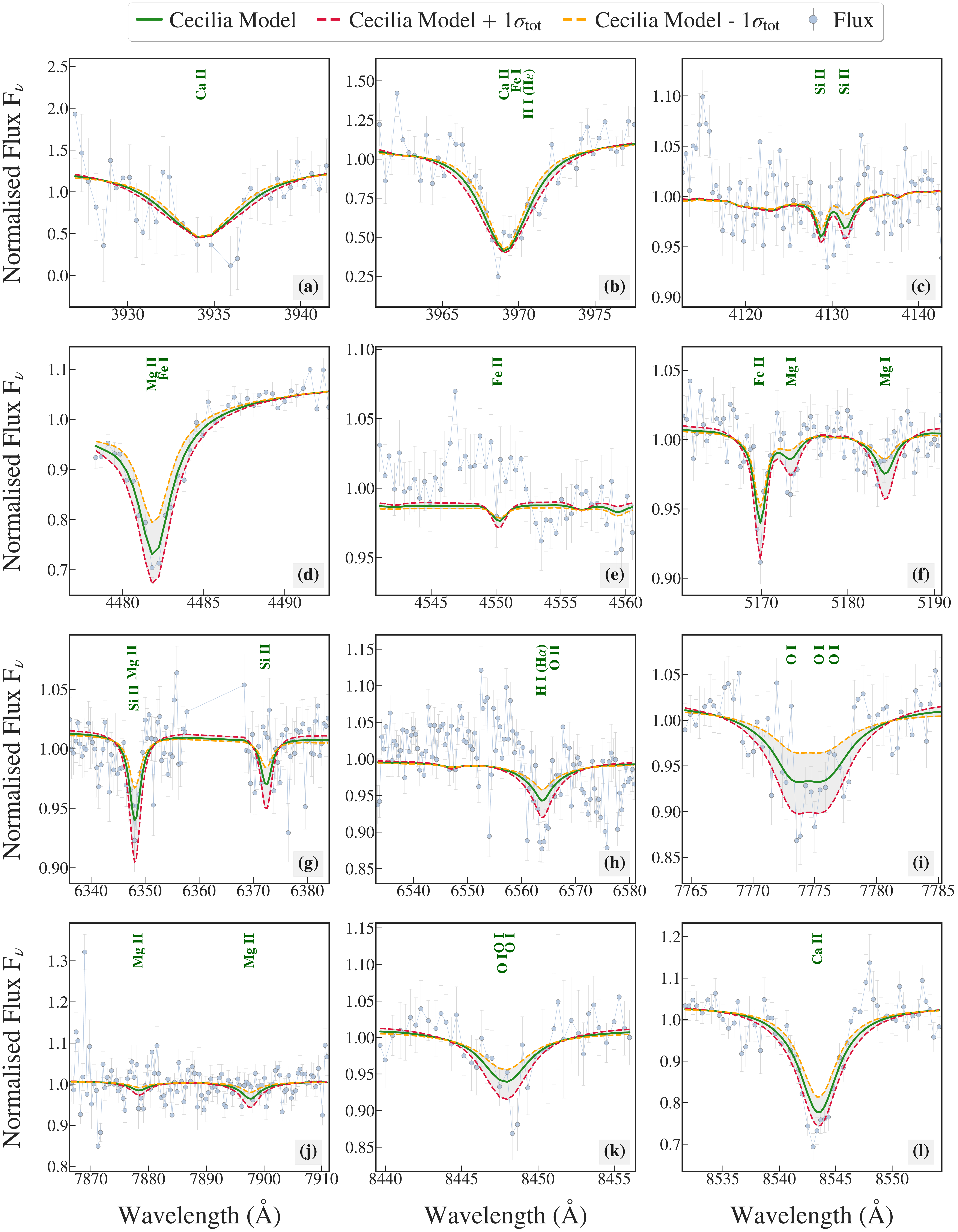}
      \caption{\cecilia's best-fit MCMC model (in green)  for the median-normalised \KeckESI{} spectrum of \wdtwothreeoneone, with its corresponding $\pm$1\sigmatot{} models (in red and orange).}
      \label{fig:paper3_fit_wd2311}
\end{figure*}

\begin{figure*}
      \centering
      \includegraphics[width=0.53\linewidth]{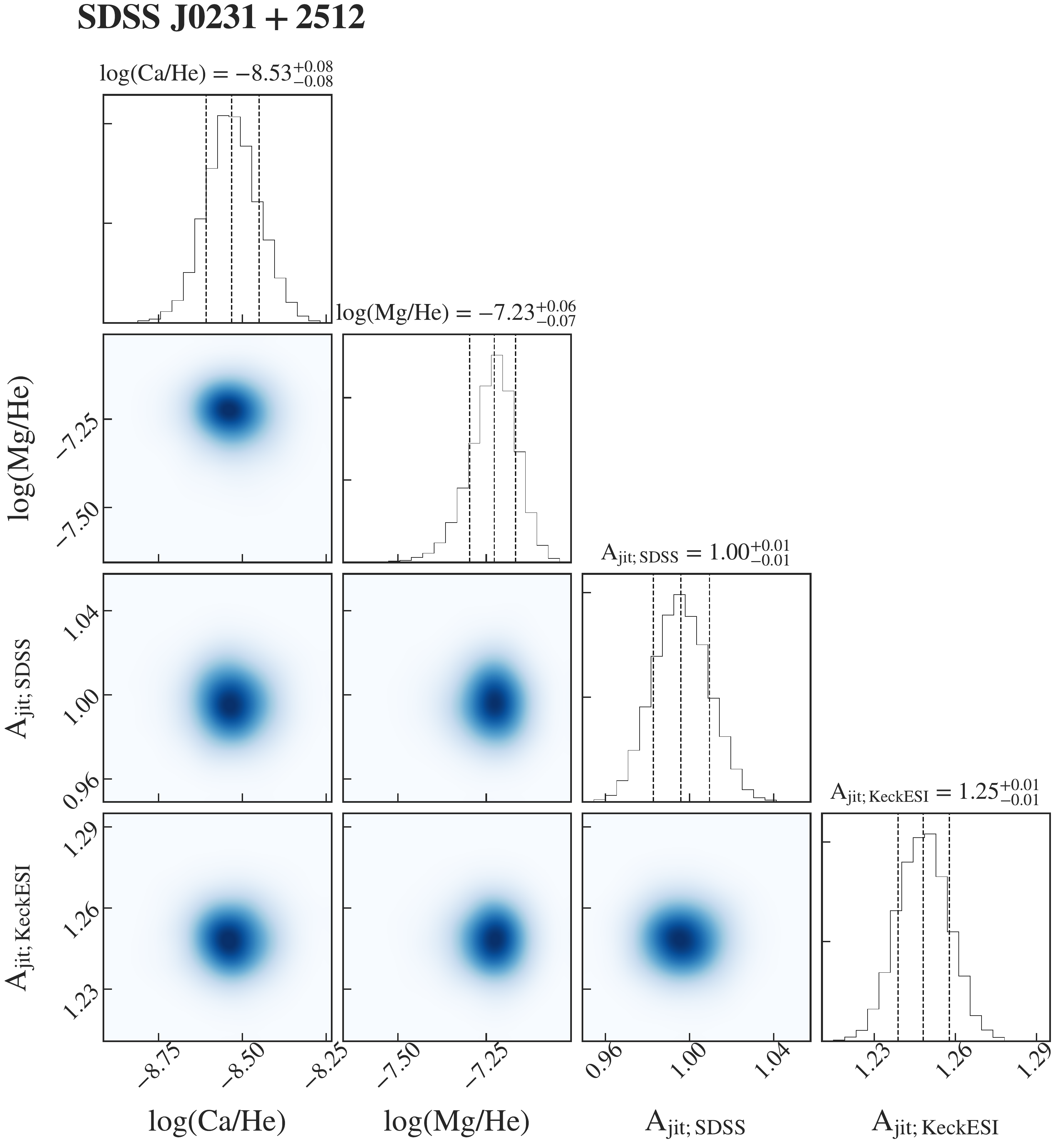}
      \caption{MCMC corner plot for \wdzerotwothreeone. }
      \label{fig:paper3_corner_wd0231}
\end{figure*}

\begin{figure*}
      \centering
      \includegraphics[width=0.6\linewidth]{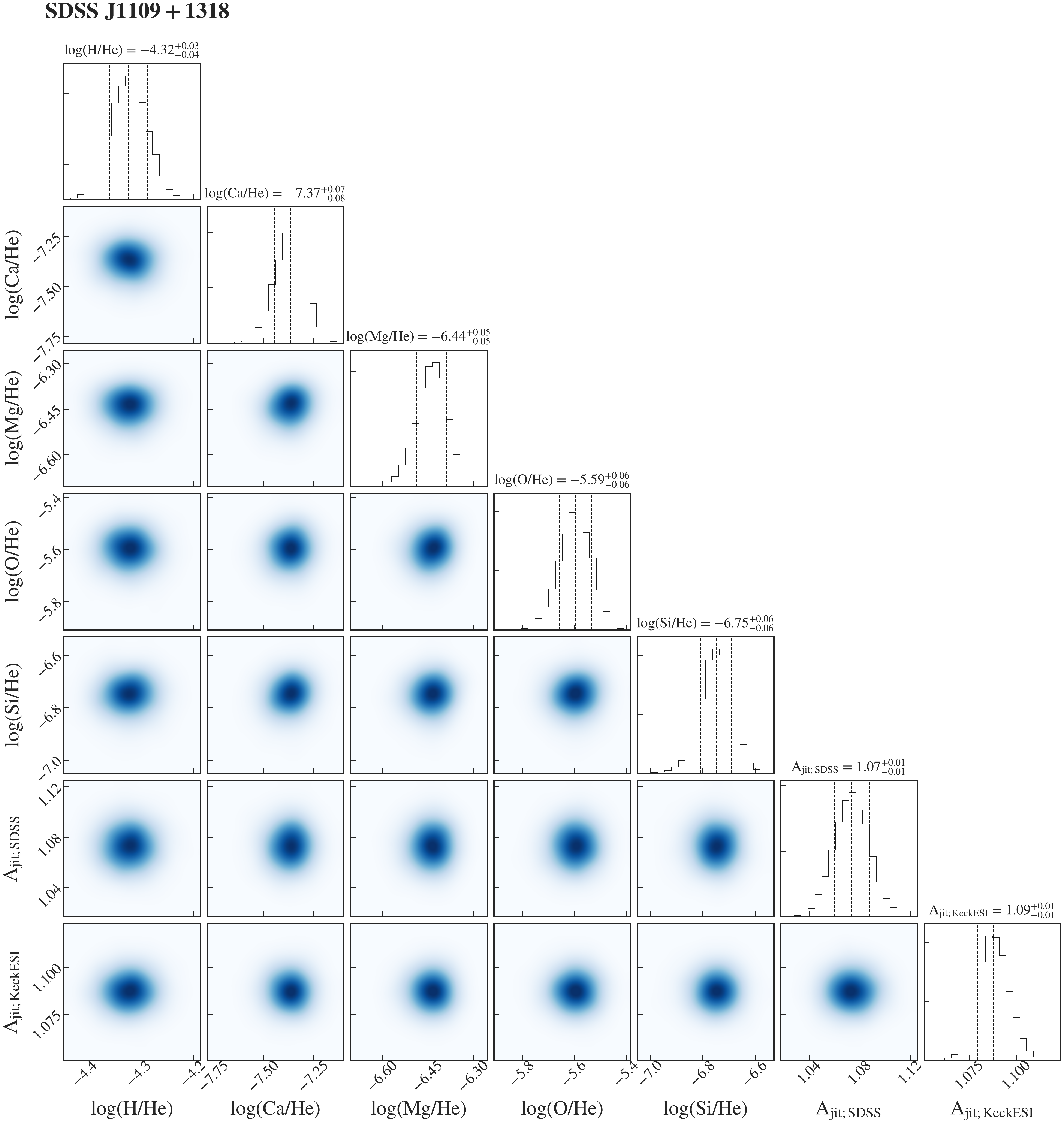}
      \caption{MCMC corner plot for \wdoneonezeronine.}
      \label{fig:paper3_corner_wd1109}
\end{figure*}

\begin{figure*}
      \centering
      \includegraphics[width=0.55\linewidth]{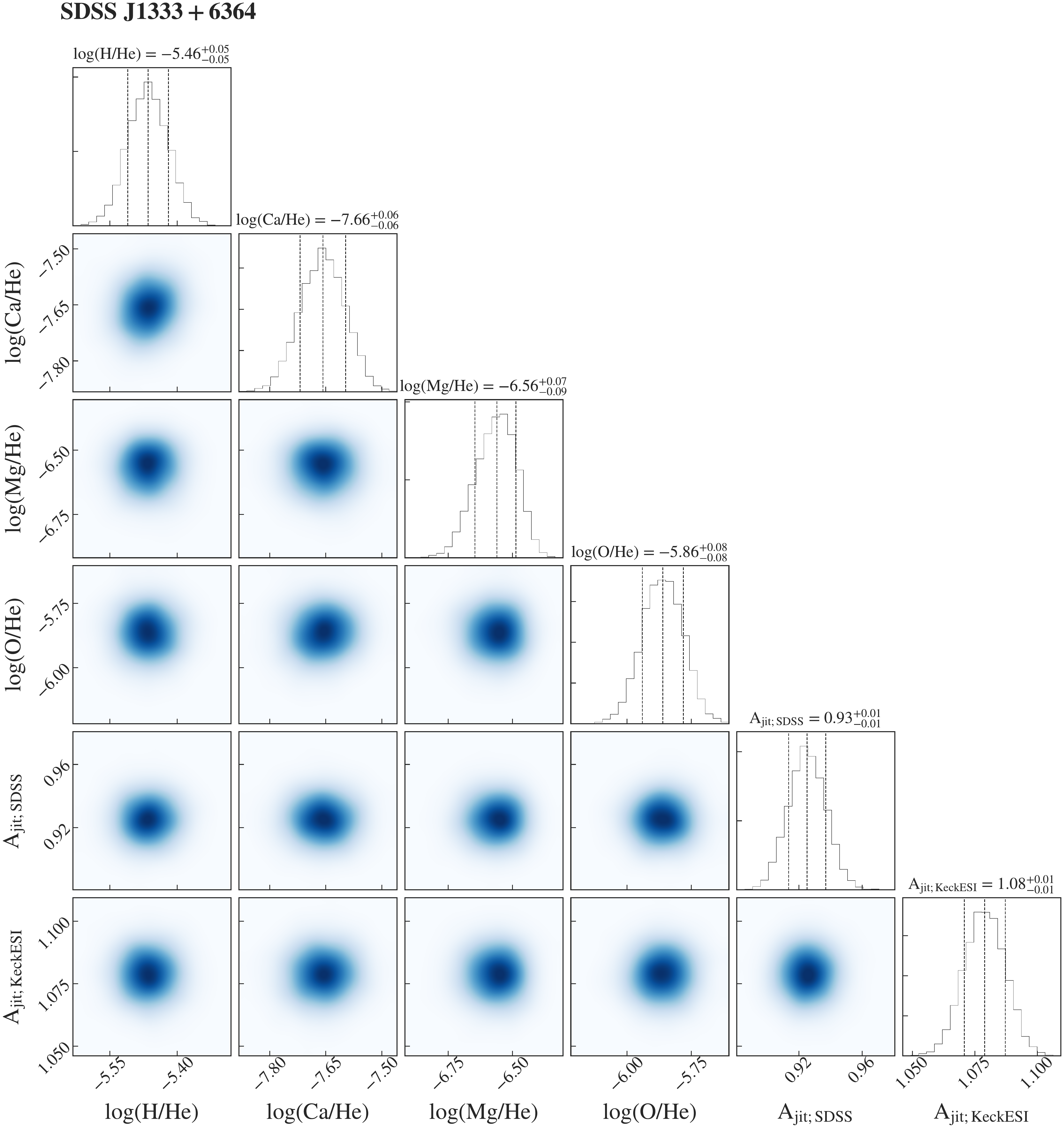}
      \caption{MCMC corner plot for \wdonethreethreethree.}
      \label{fig:paper3_corner_wd1333}
\end{figure*}

\begin{figure*}
      \centering
      \includegraphics[width=0.58\linewidth]{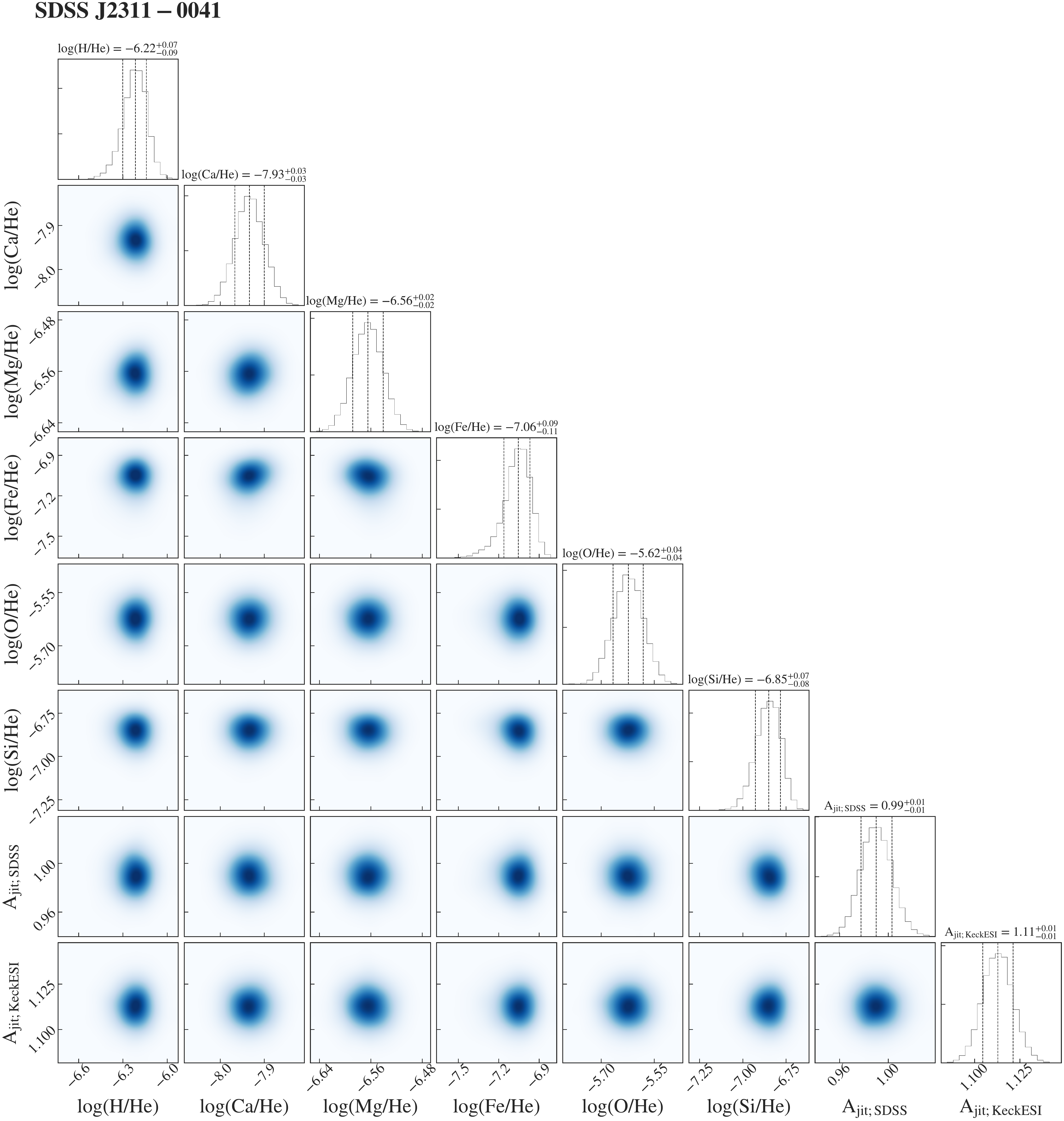}
      \caption{MCMC corner plot for \wdtwothreeoneone.}
      \label{fig:paper3_corner_wd2311}
\end{figure*}

\begin{figure*}
    \centering
    \includegraphics[width=1\linewidth]{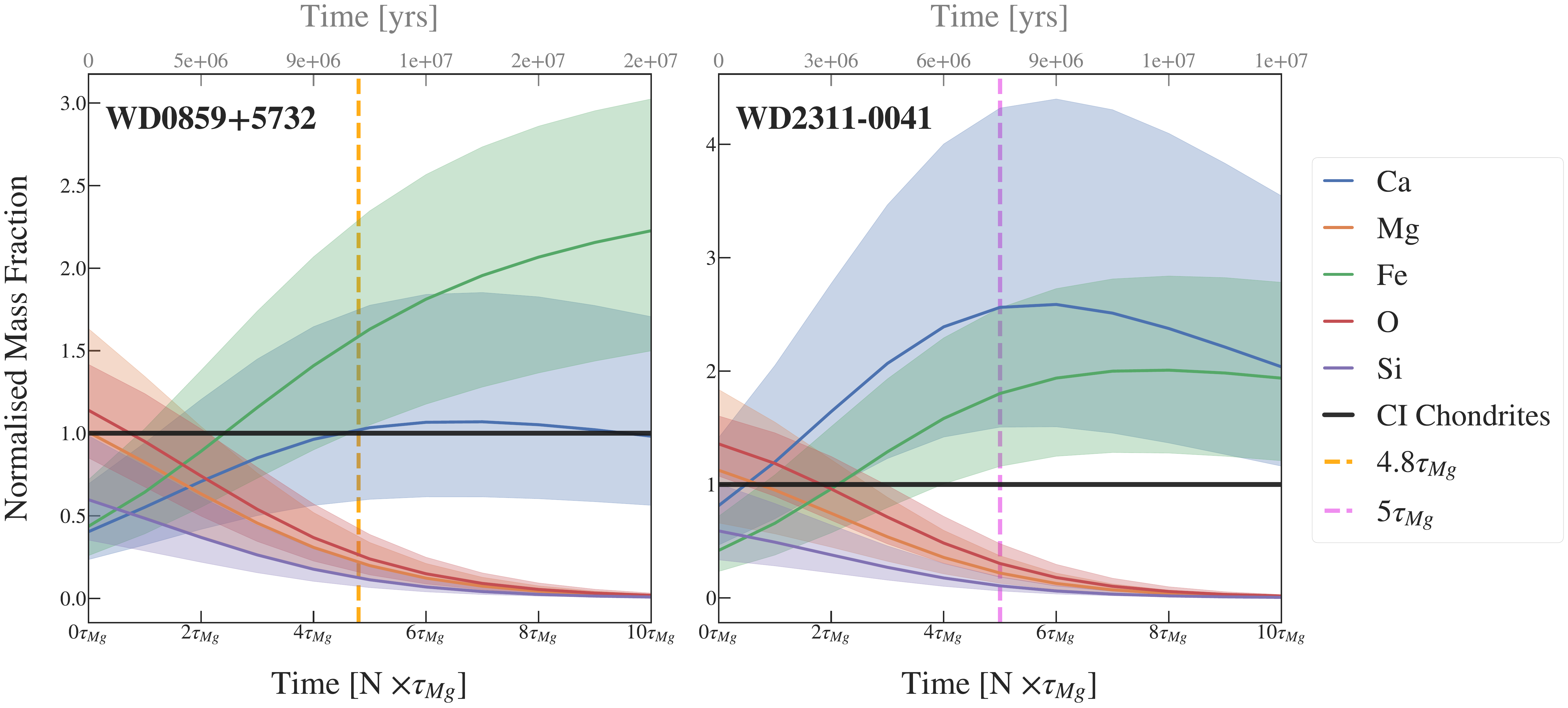}
    \caption{Mass fractions of \cecilia's detected metals for \wdzeroeightfivenine{} (left) and \wdtwothreeoneone{} (right), normalised to those of CI chondrites (black line; \citealt{Alexander:2019a_noncarbonaceous, Alexander:2019b_carbonaceous}). The dashed vertical lines at 4.8$\tau_{\mathrm{Mg}}$ and 5$\tau_{\mathrm{Mg}}$ indicate the point at which these systems cease to show evidence of oxygen excess, respectively (see Section \ref{sec:paper3_discussion_limitations}).}
   \label{fig:paper3_decay_evolution}
\end{figure*}

In Figures \ref{fig:paper3_fit_wd0231}-\ref{fig:paper3_fit_wd2311}, we present \cecilia's RV-shifted MCMC solutions (in green) for the median-normalised \KeckESI{} spectra of \wdzerotwothreeone, \wdoneonezeronine, \wdonethreethreethree, and \wdtwothreeoneone{} (in blue; air wavelengths). These figures also illustrate \cecilia's predictions when modifying the best-fit abundances of the detected elements by $\pm$1\sigmatot{} (in red and orange). For reference, we use green labels to denote \cecilia's detections, i.e. those elements with \sigmastat{}$\leq$\sigmathresh{} and at least one visible absorption line in the spectrum. In Figures \ref{fig:paper3_corner_wd0231}-\ref{fig:paper3_corner_wd2311}, we show the corner plots of \cecilia's MCMC solutions, excluding the undetected or tentative elements, as well as the the RV shifts of each spectrum. Lastly, \autoref{fig:paper3_decay_evolution} shows the continuous decay evolution of metal mass fractions for \wdzeroeightfivenine{} and \wdtwothreeoneone{}, focusing on \cecilia's detected metals.

\section{Tables}

In Table \ref{tab:paper3_table_props_detected_metals}, we provide the main compositional properties of our targets during the build-up and steady-state phases of accretion. In Table \ref{tab:paper3_classical_results}, we compare \cecilia's best-fit elemental abundances to those obtained with the classical, line-by-line fitting method of DF12 (see Section~\ref{sec:paper3_spectral_modelling} for a brief summary of the DF12 approach).

\begin{table*} 
    \centering
    \caption{Compositional properties of the five polluted WDs and their accreted material during build-up and steady-state (see Sections \ref{sec:paper3_estimation_wd_abundances}-\ref{sec:paper3_estimation_pb_abundances}). The metal mass fractions are only reported for those systems which exhibit the four main-rock forming elements at the same time (Mg, Si, Fe, and O). In this Table, we only show \cecilia's detected metals, sorted by increasing condensation temperature $T_{\rm cond}$ \citep{Lodders:2003}.}
    \label{tab:paper3_table_props_detected_metals}
    \renewcommand{\arraystretch}{1.18}
    \addtolength{\tabcolsep}{1pt}  
    \begin{tabular}{|l|c|c|c|c|c|}
     \hline 
     Property                                                 & {\bf \wdzerotwothreeone}      & {\bf \wdzeroeightfivenine}    & {\bf \wdoneonezeronine}       & {\bf \wdonethreethreethree}   & {\bf \wdtwothreeoneone}      \\
     \hline
     %%%%%%%%%%%%%%%%%%%%%%%%%%%%%%%%%%%%%%%%%%%%%%%%%%%%%%%%%%%%%%%%%%%%%%%%
     % ::: Oxygen
     %%%%%%%%%%%%%%%%%%%%%%%%%%%%%%%%%%%%%%%%%%%%%%%%%%%%%%%%%%%%%%%%%%%%%%%%
     \hline \multicolumn{6}{|c|}{\it Oxygen (T$_{\rm cond}$=182 K)} \\ \hline
     $\log_{10}(\tau_{\rm O})$ [yrs]                          & \multirow{9}{*}{Not Detected} & 6.35                          & 5.74                          & 6.07                          & 6.16                         \\
     $M_{\rm O}$ [$10^{20}~\rm{g}$]                           &                               & 1303.56$_{-485.76}^{+770.35}$ & 156.85$_{-60.10}^{+96.50}$    & 172.14$_{-67.28}^{+110.23}$   & 461.60$_{-173.57}^{+275.90}$ \\
     $\dot{M}_{\rm O}$ [$10^{8}~\rm{g/s}$; Steady]            &                               & 18.38$_{-6.85}^{+10.86}$      & 9.15$_{-3.51}^{+5.63}$        & 4.66$_{-1.82}^{+2.98}$        & 10.15$_{-3.82}^{+6.07}$      \\
     $\rm n_{O}/n_{Mg}$ [Observed; Build-up]                  &                               & 8.15$_{-3.05}^{+4.86}$        & 6.98$_{-2.71}^{+4.43}$        & 5.06$_{-2.07}^{+3.53}$        & 8.76$_{-3.31}^{+5.27}$       \\
     $\rm n_{O}/n_{Mg}$ [Steady]                              &                               & 8.27$_{-3.10}^{+4.92}$        & 7.02$_{-2.73}^{+4.46}$        & 5.06$_{-2.07}^{+3.53}$        & 8.99$_{-3.40}^{+5.41}$       \\
     $\rm MR_{O, Mg}$ [Observed; Build-up]                    &                               & 5.36$_{-2.59}^{+4.97}$        & 4.58$_{-2.24}^{+4.41}$        & 3.33$_{-1.68}^{+3.37}$        & 5.75$_{-2.78}^{+5.39}$       \\
     $\rm MR_{O, Mg}$ [Steady]                                &                               & 5.43$_{-2.62}^{+5.04}$        & 4.61$_{-2.26}^{+4.44}$        & 3.33$_{-1.68}^{+3.37}$        & 5.90$_{-2.85}^{+5.53}$       \\
     $\rm MF_{O}$ [Observed; Build-up]                        &                               & 52.42$_{-13.28}^{+12.77}$     & --                            & --                            & 62.47$_{-13.05}^{+11.33}$    \\
     $\rm MF_{O}$ [Steady]                                    &                               & 45.65$_{-12.88}^{+13.28}$     & --                            & --                            & 56.72$_{-13.36}^{+12.12}$    \\
     %%%%%%%%%%%%%%%%%%%%%%%%%%%%%%%%%%%%%%%%%%%%%%%%%%%%%%%%%%%%%%%%%%%%%%%%
     % ::: Iron
     %%%%%%%%%%%%%%%%%%%%%%%%%%%%%%%%%%%%%%%%%%%%%%%%%%%%%%%%%%%%%%%%%%%%%%%%
     \hline \multicolumn{6}{|c|}{\it Iron (T$_{\rm cond}$=1357 K)} \\ \hline
     $\log_{10}(\tau_{\rm Fe})$ [yrs]                         & \multirow{9}{*}{Not Detected} & 6.16                          & \multirow{9}{*}{Not Detected} & \multirow{9}{*}{Not Detected} & 5.96                         \\
     $M_{\rm Fe}$ [$10^{20}~\rm{g}$]                          &                               & 205.07$_{-76.84}^{+123.90}$   &                               &                               & 58.19$_{-23.56}^{+38.74}$    \\
     $\dot{M}_{\rm Fe}$ [$10^{8}~\rm{g/s}$; Steady]           &                               & 4.51$_{-1.69}^{+2.72}$        &                               &                               & 2.01$_{-0.81}^{+1.34}$       \\
     $\rm n_{Fe}/n_{Mg}$ [Observed; Build-up]                 &                               & 0.37$_{-0.14}^{+0.22}$        &                               &                               & 0.32$_{-0.13}^{+0.21}$       \\
     $\rm n_{Fe}/n_{Mg}$ [Steady]                             &                               & 0.58$_{-0.22}^{+0.35}$        &                               &                               & 0.51$_{-0.21}^{+0.34}$       \\
     $\rm MR_{Fe, Mg}$ [Observed; Build-up]                   &                               & 0.84$_{-0.41}^{+0.78}$        &                               &                               & 0.72$_{-0.36}^{+0.72}$       \\
     $\rm MR_{Fe, Mg}$ [Steady]                               &                               & 1.33$_{-0.64}^{+1.24}$        &                               &                               & 1.16$_{-0.59}^{+1.16}$       \\
     $\rm MF_{Fe}$ [Observed; Build-up]                       &                               & 8.07$_{-3.28}^{+5.22}$        &                               &                               & 7.75$_{-3.43}^{+5.57}$       \\
     $\rm MF_{Fe}$ [Steady]                                   &                               & 10.98$_{-4.34}^{+6.69}$       &                               &                               & 11.07$_{-4.77}^{+7.36}$      \\
     %%%%%%%%%%%%%%%%%%%%%%%%%%%%%%%%%%%%%%%%%%%%%%%%%%%%%%%%%%%%%%%%%%%%%%%%
     % ::: Magnesium
     %%%%%%%%%%%%%%%%%%%%%%%%%%%%%%%%%%%%%%%%%%%%%%%%%%%%%%%%%%%%%%%%%%%%%%%%
     \hline \multicolumn{6}{|c|}{\it Magnesium (T$_{\rm cond}$=1397 K)} \\ \hline
     $\log_{10}(\tau_{\rm Mg})$ [yrs]                         & 6.72                          & 6.36                          & 5.74                          & 6.07                          & 6.17                         \\
     $M_{\rm Mg}$ [$10^{20}~\rm{g}$]                          & 50.19$_{-19.37}^{+31.39}$     & 243.31$_{-90.05}^{+143.83}$   & 34.23$_{-12.91}^{+20.68}$     & 51.70$_{-20.14}^{+32.94}$     & 80.17$_{-29.67}^{+47.46}$    \\
     $\dot{M}_{\rm Mg}$ [$10^{8}~\rm{g/s}$; Steady]           & 0.30$_{-0.12}^{+0.19}$        & 3.38$_{-1.25}^{+2.00}$        & 1.98$_{-0.75}^{+1.20}$        & 1.40$_{-0.54}^{+0.89}$        & 1.72$_{-0.64}^{+1.02}$       \\
     $\rm n_{Mg}/n_{Mg}$ [Observed; Build-up]                 & 1.00$_{-0.37}^{+0.59}$        & 1.00$_{-0.37}^{+0.59}$        & 1.00$_{-0.37}^{+0.59}$        & 1.00$_{-0.37}^{+0.59}$        & 1.00$_{-0.37}^{+0.59}$       \\
     $\rm n_{Mg}/n_{Mg}$ [Steady]                             & 1.00$_{-0.37}^{+0.59}$        & 1.00$_{-0.37}^{+0.59}$        & 1.00$_{-0.37}^{+0.59}$        & 1.00$_{-0.37}^{+0.59}$        & 1.00$_{-0.37}^{+0.59}$       \\
     $\rm MR_{Mg, Mg}$ [Observed; Build-up]                   & 1.00$_{-0.00}^{+0.00}$        & 1.00$_{-0.00}^{+0.00}$        & 1.00$_{-0.00}^{+0.00}$        & 1.00$_{-0.00}^{+0.00}$        & 1.00$_{-0.00}^{+0.00}$      \\
     $\rm MR_{Mg, Mg}$ [Steady]                               & 1.00$_{-0.00}^{+0.00}$        & 1.00$_{-0.00}^{+0.00}$        & 1.00$_{-0.00}^{+0.00}$        & 1.00$_{-0.00}^{+0.00}$        & 1.00$_{-0.00}^{+0.00}$       \\
     $\rm MF_{Mg}$ [Observed; Build-up]                       & --                            & 9.60$_{-3.87}^{+5.98}$        & --                            & --                            & 10.74$_{-4.43}^{+6.79}$      \\
     $\rm MF_{Mg}$ [Steady]                                   & --                            & 8.22$_{-3.29}^{+5.17}$        & --                            & --                            & 9.45$_{-3.85}^{+5.96}$       \\
     %%%%%%%%%%%%%%%%%%%%%%%%%%%%%%%%%%%%%%%%%%%%%%%%%%%%%%%%%%%%%%%%%%%%%%%%
     % ::: Silicon
     %%%%%%%%%%%%%%%%%%%%%%%%%%%%%%%%%%%%%%%%%%%%%%%%%%%%%%%%%%%%%%%%%%%%%%%%
     \hline \multicolumn{6}{|c|}{\it Silicon (T$_{\rm cond}$=1529 K)} \\ \hline
     $\log_{10}(\tau_{\rm Si})$ [yrs]                         & \multirow{9}{*}{Not Detected} & 6.36                          & 5.72                          & \multirow{9}{*}{Not Detected} & 6.18                         \\
     $M_{\rm Si}$ [$10^{20}~\rm{g}$]                          &                               & 162.48$_{-60.31}^{+96.50}$    & 19.33$_{-7.35}^{+11.95}$      &                               & 47.44$_{-18.27}^{+29.85}$    \\
     $\dot{M}_{\rm Si}$ [$10^{8}~\rm{g/s}$; Steady]           &                               & 2.24$_{-0.83}^{+1.33}$        & 1.17$_{-0.44}^{+0.72}$        &                               & 1.00$_{-0.39}^{+0.63}$       \\
     $\rm n_{Si}/n_{Mg}$ [Observed; Build-up]                 &                               & 0.58$_{-0.21}^{+0.35}$        & 0.49$_{-0.19}^{+0.31}$        &                               & 0.51$_{-0.20}^{+0.32}$       \\
     $\rm n_{Si}/n_{Mg}$ [Steady]                             &                               & 0.57$_{-0.21}^{+0.34}$        & 0.51$_{-0.20}^{+0.32}$        &                               & 0.50$_{-0.20}^{+0.32}$       \\
     $\rm MR_{Si, Mg}$ [Observed; Build-up]                   &                               & 0.67$_{-0.32}^{+0.61}$        & 0.57$_{-0.28}^{+0.54}$        &                               & 0.59$_{-0.29}^{+0.56}$       \\
     $\rm MR_{Si, Mg}$ [Steady]                               &                               & 0.66$_{-0.32}^{+0.61}$        & 0.59$_{-0.29}^{+0.56}$        &                               & 0.58$_{-0.29}^{+0.55}$       \\
     $\rm MF_{Si}$ [Observed; Build-up]                       &                               & 6.39$_{-2.62}^{+4.14}$        & --                            &                               & 6.32$_{-2.73}^{+4.44}$       \\
     $\rm MF_{Si}$ [Steady]                                   &                               & 5.42$_{-2.21}^{+3.53}$        & --                            &                               & 5.48$_{-2.34}^{+3.82}$       \\
     %%%%%%%%%%%%%%%%%%%%%%%%%%%%%%%%%%%%%%%%%%%%%%%%%%%%%%%%%%%%%%%%%%%%%%%%
     % ::: Calcium
     %%%%%%%%%%%%%%%%%%%%%%%%%%%%%%%%%%%%%%%%%%%%%%%%%%%%%%%%%%%%%%%%%%%%%%%%
     \hline \multicolumn{6}{|c|}{\it Calcium (T$_{\rm cond}$=1659 K)} \\ \hline
     $\log_{10}(\tau_{\rm Ca})$ [yrs]                         & 6.55                          & 6.18                          & 5.55                          & 5.89                          & 5.98                         \\
     $M_{\rm Ca}$ [$10^{20}~\rm{g}$]                          & 4.15$_{-1.62}^{+2.65}$        & 9.44$_{-3.51}^{+5.71}$        & 6.60$_{-2.55}^{+4.23}$        & 6.84$_{-2.60}^{+4.26}$        & 5.62$_{-2.09}^{+3.36}$       \\
     $\dot{M}_{\rm Ca}$ [$10^{8}~\rm{g/s}$; Steady]           & 0.04$_{-0.01}^{+0.02}$        & 0.20$_{-0.07}^{+0.12}$        & 0.59$_{-0.23}^{+0.38}$        & 0.28$_{-0.11}^{+0.17}$        & 0.19$_{-0.07}^{+0.11}$       \\
     $\rm n_{Ca}/n_{Mg}$ [Observed; Build-up]                 & 0.05$_{-0.02}^{+0.03}$        & 0.02$_{-0.01}^{+0.01}$        & 0.12$_{-0.05}^{+0.08}$        & 0.08$_{-0.03}^{+0.05}$        & 0.04$_{-0.02}^{+0.03}$       \\
     $\rm n_{Ca}/n_{Mg}$ [Steady]                             & 0.07$_{-0.03}^{+0.05}$        & 0.04$_{-0.01}^{+0.02}$        & 0.18$_{-0.07}^{+0.12}$        & 0.12$_{-0.05}^{+0.08}$        & 0.07$_{-0.02}^{+0.04}$       \\
     $\rm MR_{Ca, Mg}$ [Observed; Build-up]                   & 0.08$_{-0.04}^{+0.08}$        & 0.04$_{-0.02}^{+0.04}$        & 0.19$_{-0.09}^{+0.19}$        & 0.13$_{-0.07}^{+0.13}$        & 0.07$_{-0.03}^{+0.07}$       \\
     $\rm MR_{Ca, Mg}$ [Steady]                               & 0.12$_{-0.06}^{+0.12}$        & 0.06$_{-0.03}^{+0.06}$        & 0.30$_{-0.15}^{+0.29}$        & 0.20$_{-0.10}^{+0.20}$        & 0.11$_{-0.05}^{+0.10}$       \\
     $\rm MF_{Ca}$ [Observed; Build-up]                       & --                            & 0.37$_{-0.15}^{+0.26}$        & --                            & --                            & 0.74$_{-0.32}^{+0.55}$       \\
     $\rm MF_{Ca}$ [Steady]                                   & --                            & 0.48$_{-0.20}^{+0.34}$        & --                            & --                            & 1.02$_{-0.43}^{+0.73}$       \\
     \hline \hline
     \hfill  \Mcvz{} [$10^{-6}$~\Msun]                        & 7.08                          & 3.51                          & 0.78                          & 1.57                          & 2.43                         \\
     \hfill $\boldsymbol{\sum M_{\rm Z}}$ [g]                 & $>5.43\times10^{21}$          & $>1.92\times10^{23}$          & $>2.17\times10^{22}$          & $>2.31\times10^{22}$          & $>6.53\times10^{22}$         \\
     \hfill $\boldsymbol{\sum{\dot{M}_{\rm Z}}}$ [$\rm{g/s}$] & $>3.40\times10^{7}$           & $>2.87\times10^{9}$           & $>1.29\times10^{9}$           & $>6.33\times10^{8}$           & $>1.51\times10^{9}$          \\ \hline
    \end{tabular}
\end{table*}

\begin{table*} 
    \centering
    \caption{Comparison between the elemental abundances obtained with the classical, line-by-line fitting method of DF12 (left) and those predicted by \cecilia's ML-based optimisation routine. The parentheses in the DF12 columns denote the number of absorption lines used to constrain each abundance measurement. The uncertainties of \cecilia's results represent the total error (\sigmatot), calculated as the quadrature sum of the statistical (\sigmastat) and systematic error sources (\sigmasys). As discussed in Section \ref{sec:paper3_results}, \cecilia's estimated abundances are consistent with those of DF12 within 0.2~dex.} 
    \label{tab:paper3_classical_results}
    \renewcommand{\arraystretch}{1.9}
    \addtolength{\tabcolsep}{-4.3pt}  
    \begin{tabular}{|l|c|c|c|c|c|c|c|c|c|c|}
        \hline
        \multirow{2}{*}{Property} & \multicolumn{2}{c|}{{\bf \wdzerotwothreeone}} & \multicolumn{2}{c|}{{\bf \wdzeroeightfivenine}} & \multicolumn{2}{c|}{{\bf \wdoneonezeronine}} & \multicolumn{2}{c|}{{\bf \wdonethreethreethree}} & \multicolumn{2}{c|}{{\bf \wdtwothreeoneone}} \\
        \cline{2-11}
         & DF12 & \cecilia{} & DF12 & \cecilia{} & DF12 & \cecilia{} & DF12 & \cecilia{} & DF12 & \cecilia{} \\
        \hline\hline
        \logHHe{}  & -6.26 {\footnotesize(1)}          & --              & --                                & -6.60$\pm$0.20  & -4.39 (1)                         & -4.33$\pm$0.20  & -5.40 {\footnotesize(1)}                                & -5.49$\pm$0.20  & -6.07 {\footnotesize(1)}          & -6.26$\pm$0.20  \\
        \logOHe{}  & --                                & --              & -5.37$\pm$0.10 {\footnotesize(2)} & -5.35$\pm$0.20  & -5.35$\pm$0.19 {\footnotesize(2)} & -5.66$\pm$0.20  & -5.49$\pm$0.24 {\footnotesize(2)}                       & -5.85$\pm$0.20  & -5.51$\pm$0.15 {\footnotesize(2)} & -5.64$\pm$0.20  \\
        \logMgHe{} & -7.35 {\footnotesize(1)}          & -7.22$\pm$0.20  & -6.20$\pm$0.07 {\footnotesize(3)} & -6.24$\pm$0.20  & -6.57 {\footnotesize(1)}          & -6.46$\pm$0.20  & -6.48 {\footnotesize(1)}                                & -6.54$\pm$0.20  & -6.60 {\footnotesize(1)}          & -6.57$\pm$0.20  \\
        \logSiHe{} & -6.90 {\footnotesize(1)}          & --              & -6.30$\pm$0.10 {\footnotesize(3)} & -6.48$\pm$0.20  & -6.44$\pm$0.11 {\footnotesize(3)} & -6.72$\pm$0.20  & -6.91$\pm$0.16 {\footnotesize(3)}                       &   --            & -6.69$\pm$0.15 {\footnotesize(3)} & -6.78$\pm$0.20  \\
        \logCaHe{} & -8.89$\pm$0.10 {\footnotesize(2)} & -8.58$\pm$0.20  & -8.00$\pm$0.17 {\footnotesize(3)} & -7.86$\pm$0.20  & -7.58$\pm$0.17 {\footnotesize(3)} & -7.34$\pm$0.20  & -7.50$\pm$0.20 {\footnotesize(3)} & -7.78$\pm$0.20  & -7.97$\pm$0.04 {\footnotesize(3)} & -7.94$\pm$0.20  \\
        \logFeHe{} & --                                &                 & -6.62$\pm$0.13 {\footnotesize(3)} & -6.66$\pm$0.20  & --                                &                 & --                                                      &                 & -6.77$\pm$0.18 {\footnotesize(2)} & -7.09$\pm$0.20  \\
        \hline
    \end{tabular}
    \vspace{0.1cm}
    %\begin{quote}
    %    \footnotesize
    %    \hspace{-0.17cm}\footnotesize{[\it{a}}]: For \wdzerotwothreeone, calcium measurements are based solely on SDSS data as other spectra were too noisy.\\
    %    \hspace{-0.17cm}\footnotesize{[\it{b}}]: For WD1333, calcium measurements are noisy around 3,933\angs. Using SDSS data, a value of -7.5 seems more representative.
    %\end{quote}
\end{table*}
%%%%%%%%%%%%%%%%%%%%%%%%%%%%%%%%%%%%%%%%%%%%%%%%%%

% Don't change these lines
\bsp	% typesetting comment
\label{lastpage}
\end{document}